\documentclass[
aps,twocolumn,floatfix,longbibliography 
]{revtex4-2}

\usepackage[dvips]{graphicx}
\usepackage{bbm}
\usepackage{amsmath}
\usepackage{amsfonts} 
\usepackage{mathtools}
\usepackage[colorlinks=true,allcolors=blue]{hyperref}

\usepackage{braket} 

\newcommand{\sect}[1]{\vspace{0.3em}{\textit{#1.}}---}

\newcommand{\manuscriptTitle}{{Readout of multi-level quantum geometry from electronic transport}}
\newcommand{\UniA}{{Institute of Physics, University of Augsburg, D-86159 Augsburg, Germany}}

\allowdisplaybreaks

\begin{document}

\title{\manuscriptTitle}
\author{Raffael~L.~Klees}
\email{raffael.klees@uni-a.de}
\affiliation{\UniA}
\author{M{\'o}nica~Benito}
\affiliation{\UniA}

\begin{abstract}
    The quantum geometric tensor (QGT) of a quantum system in a given parameter space captures both the geometry of the state manifold and the topology of the system. While the local QGT elements have been successfully measured in various platforms, the challenge of detecting them in electronic transport systems -- such as tunnel or molecular junctions -- has yet to be resolved.
    To fill this gap, we propose a measurement protocol based on weak and resonant parameter modulations, and theoretically demonstrate how the local QGT in such systems can be directly probed from changes of the tunnel conductance. This approach enables the measurement of both geometrical and topological features of quantum states in a broad class of transport-based quantum systems.
\end{abstract} 
\date{\today}
\maketitle

\begin{figure}
    \centering
    \includegraphics{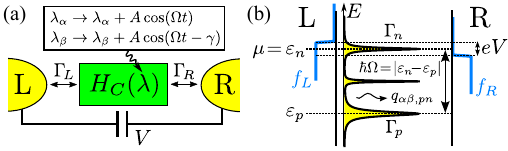} 
    \caption{
        (a) A system with Hamiltonian $H_C(\lambda)$ depending on parameters $\lambda = (\lambda_\alpha)$ is connected to left (L) and right (R) terminals with couplings $\Gamma_{L}$ and $\Gamma_R$, respectively, and a voltage is applied.
        The center parameters $\lambda_\alpha$ are driven with amplitude $A$ at frequency $\Omega$.
        A simultaneous drive of two parameters $\lambda_\alpha$ and $\lambda_\beta$ may involve a relative phase $\gamma$. 
        (b) The energies $\varepsilon_n$ of the center gain a spectral broadening $\Gamma_n$.
        The linear conductance through an energy level $\varepsilon_n$ is measured by shifting the chemical potential $\mu = \varepsilon_n$ in both contacts and applying a small relative voltage $V$. 
        Driving the parameters resonantly at frequency $\hbar \Omega = |\varepsilon_n - \varepsilon_p|$ leads to a change in conductance proportional to the QGT transition elements $q_{\alpha\beta,pn}$ defined in Eq.~\eqref{eq:qgt}.
        Different linear and circular driving schemes, enabled by special choices of $\gamma$ in panel (a), allow to gain access to the different $(\alpha,\beta)$ elements. 
    }
    \label{fig:fig1}
\end{figure}

%
%
\sect{Introduction}The quantum geometric tensor (QGT) is a fundamental object in the study of quantum systems and reflects the internal geometry of quantum states in an arbitrary parameter space. 
It naturally arises in different branches of physics and provides a unified framework for understanding how the quantum states and topology of a system change with changing these parameters \cite{Provost1980,Resta2011,Kolodrubetz2013,Kolodrubetz2017,Torma2023}. 
For instance, geometric concepts related with the QGT appear in topological quantum materials \cite{Hasan2010,Sato2017,Armitage2018,Liu2024,Yu2025}, affect superfluidity and superconductivity in flat-band systems \cite{Peotta2015,Liang2017,Torma2018,Torma2022,Tian2023,Peotta2023}, are relevant for holonomic quantum gates \cite{Zanardi1999,Toyoda2013,Leroux2018,Yang2019,Sugawa2021} and quantum computing \cite{Viyuela2018,Bleu2018b,yamamoto2019,McArdle2019,Stokes2020,Cerezo2021,Koczor2022,Chen2024}, and contribute to nonlinear photoconductivity effects \cite{Ahn2020,Hsu2023,Jankowski2024,Jankowski2024b,Avdoshkin2025}.
These concepts have been generalized in several directions, including non-Abelian \cite{Ma2010,Zheng2022,Ding2024,mitscherling2025} and mixed quantum states \cite{Dittmann1992,Viyuela2014,Gneiting2022,Hou2024,Zhou2024,Wang2025,Wang2025b,Wang2025c,ji2025}, non-Hermitian and open quantum systems \cite{Zhang2019,Farias2021,Bergholtz2021,ChenYe2024,Orlov2025,Hu2024,Behrends2025,Hu2025}, tensor monopoles \cite{Palumbo2018,Palumbo2019,Weisbrich2021b,Tan2021,jankowski2025}, and Plücker embeddings \cite{bouhon2023}.
From the experimental side, geometric and topological properties of quantum states have been measured in a variety of platforms \cite{Flaeschner2016,Sugawa2018,Bleu2018,Tan2018,Asteria2019,Tan2019,Yu2019,Gianfrate2020,Chen2022,Yu2022,Yi2023,Kim2025,Kang2025,Pai2025}.
However, to this date, the measurement of the QGT in general electronic transport setups, sketched in Fig.~\ref{fig:fig1}(a), remains an open question.

In this Letter, we present a general measurement protocol based on the electronic conductance through the device, which allows the reconstruction of the full quantum geometry of the states in the center. 
The whole idea of our proposal is summarized schematically in Fig.~\ref{fig:fig1}.
For this, we revisit the concept of parametric pumping \cite{Kouwenhoven1991,Pothier1992,Bruder1994,Spivak1995,Stoof1996,Oosterkamp1997,Brune1997,Flensberg1997,Pedersen1998,Aleiner1998,Brouwer1998,Wang2002,Kohler2005,Thingna2014,Ridley2017} and combine it with engineered weak and resonant drives of the parameters of the central system \cite{Ozawa2018}. 
The generated changes in conductance are directly linked with the elements of the local QGT.
	
\sect{Quantum geometry and transport setup}The electronic transport setup, schematically sketched in Fig.~\ref{fig:fig1}(a), is described by the Hamiltonian $H = H_L + H_R + H_C(\lambda) + H_{LC} + H_{CR}$. 
Here, $H_a$ describes normal-metal contact $a \in \{L,R\}$, $H_{LC}$ and $H_{CR}$ describe the couplings of the center to said contacts. 
$H_C(\lambda)$ describes the center of which we want to read out the QGT defined in parameter space $\lambda =(\lambda_\alpha)$.

We assume that the center hosts $N$ nondegenerate energy levels $\varepsilon_n(\lambda)$ ($n = 1,\ldots,N$) such that $H_C(\lambda)\ket{n(\lambda)} = \varepsilon_n(\lambda) \ket{n(\lambda)}$.
The gauge-invariant, hermitian QGT of an eigenstate $\ket{n}$ is $q_{\alpha\beta,n} = \bra{\partial_\alpha n} (\mathbbm{1} - \ket{n}\bra{n}) \ket{\partial_\beta n}$, with $\partial_\alpha = \partial / \partial \lambda_\alpha$ \cite{Kolodrubetz2013,Kolodrubetz2017}. 
It combines the (Fubini-Study) metric $g_{\alpha\beta,n} = \mathrm{Re}(q_{\alpha\beta,n})$ and the Berry curvature $F_{\alpha\beta,n} = -2 \, \mathrm{Im}(q_{\alpha\beta,n})$ in a single quantity.
The former defines a notion of distance $ds^2 = \sum_{\alpha\beta} g_{\alpha\beta,n} d\lambda_\alpha d\lambda_\beta$ between nearby quantum states $\ket{n(\lambda)}$ and $\ket{n(\lambda + d\lambda)}$ for small $d\lambda$, which is related with the state fidelity.
The latter contains topological information of the quantum states such as their Chern number, which follows from a suitable integration of the Berry curvature $F_{\alpha\beta,n}$ over the parameter space.
Since $F_{\alpha\alpha,n} = 0$, the diagonal elements of the QGT, $q_{\alpha\alpha,n} = g_{\alpha\alpha,n}$, are strictly real.
A useful decomposition of the QGT into transition elements $q_{\alpha\beta,pn} = q_{\beta\alpha,pn}^* = q_{\alpha\beta,np}^*$ reads
\begin{align}
    \label{eq:qgt}
    q_{\alpha\beta,n} = 
    \sum_{p=1 \atop p\neq n}^N  \braket{\partial_\alpha n|p} \braket{p|\partial_\beta n}
    \equiv
    \sum_{p=1 \atop p\neq n}^N 
    q_{\alpha\beta,pn}.
\end{align}
    
\sect{Current through a driven center}To access the local geometry of the underlying state manifold, it is not enough to perform the conductance measurement at a static parameter value $\lambda$. 
However, as we show in the following, periodically modulating the parameters $\lambda \to \lambda(t)$ around a fixed value during the measurement provides a way to probe the local curvature and allows for the reconstruction of the full QGT $q_{\alpha\beta,n}$. 

We consider specific one- and two-parameter drives in the spirit of Ref.~\cite{Ozawa2018}, which was subsequently applied in both theory \cite{Ozawa2019,Klees2020,Klees2021,Weisbrich2021} and experiment \cite{Flaeschner2016,Asteria2019,Yu2022} across different platforms. 
As sketched in Fig.~\ref{fig:fig1}, resonantly driving one parameter $\lambda_\alpha \to \lambda_\alpha + A \cos(\Omega t)$ at a frequency $\Omega$ with amplitude $A$ will give access to the diagonal transition elements $q_{\alpha\alpha,pn}$.
A simultaneous drive of two parameters, $(\lambda_\alpha,\lambda_\beta) \to (\lambda_\alpha + A \cos(\Omega t), \lambda_\beta + A \cos(\Omega t-\gamma))$ for $\alpha \neq \beta$, together with specific combinations of conductance measurements taken at different values of the relative phase difference $\gamma$, will give access to the off-diagonal transition elements $q_{\alpha\beta,pn}$. 

The Hamiltonian of the driven center becomes time-dependent, $H_C(\lambda) \to H(t) \equiv H_C(\lambda(t)) = H(t + T)$ with period $T = 2\pi/\Omega$.
To access the QGT of the bare undriven center, it is useful to extract the original center Hamiltonian $H_C$ by writing $H(t) = H_C + W(t)$, where the drive is contained in the $T$-periodic perturbation $W(t) = \sum_{k\in\mathbb{Z}} W_k e^{ik\Omega t}$ with Fourier coefficients 
\begin{align}
    \label{eq:FourierCoefficientsMainText}
    W_k  = \frac{1}{T} \int_0^T H(t)  e^{-ik\Omega t} \, dt - \delta_{k0} H_C .
\end{align}

We employ the nonequilibrium Keldysh-Green's function formalism to determine the conductance through the driven center.
Utilizing the $T$-periodicity, the resulting current is expressed as $I(t) = \sum_{k \in \mathbb{Z}} I_k e^{ik\Omega t}$.
The general relations to determine all components $I_k$ are derived in detail in the Supplemental Material (SM) \cite{supmat}.
To access the QGT it is enough to consider the time-averaged current given by the dc component $I_0 = I_\mathrm{pump} + I_\mathrm{Landauer}$, which generally splits into pump and Landauer contributions \cite{Kohler2005,supmat},
\begin{align} 
\label{eq:averageCurrent}
\begin{split}
    I_\mathrm{pump}
    &= 
    \frac{e}{h}  
    \int_\mathbb{R} 
    \sum_{k \in \mathbb{Z}}     
    \mathcal{T}^{(k)}_{\mathrm{rel}}(E)
    [ f_L(E) + f_R(E)]
    dE ,
    \\
    I_\mathrm{Landauer}
    &= 
    \frac{e}{h}  
    \int_\mathbb{R} 
    \sum_{k \in \mathbb{Z}}     
    \mathcal{T}^{(k)}_{\mathrm{avg}}(E)
    [ f_L(E)  - f_R(E)]
    dE  ,
\end{split} 
\end{align}
with the electron charge $e$ and the Planck constant $h$, respectively.
Here, $\mathcal{T}^{(k)}_{\mathrm{avg,rel}} = 	(\mathcal{T}^{(k)}_{RL} \pm \mathcal{T}^{(k)}_{LR})/2$ with $\mathcal{T}_{ab}^{(k)}(E) =
\mathrm{Tr}( \Gamma_a \mathcal{G}_{k}(E) 
\Gamma_b [\mathcal{G}_{k}(E)]^\dag)$ are the average and relative transmission functions, and 
$f_{a}(E) = [1 + e^{\beta(E-\mu_a)}]^{-1}$ is the Fermi function in contact $a$ with inverse temperature $\beta$ and chemical potential $\mu_a$.
At zero voltage, $\mu_a \equiv \mu$ and $I_0 = I_\mathrm{pump}$ becomes a pure pump current due to the drive.
Without the drive, $A = 0$, we have $\mathcal{T}_{ab}^{(k)} = 0$ for $k \neq 0$ and $\mathcal{T}_{ab}^{(0)} = \mathcal{T}_{ba}^{(0)}$, so $\mathcal{T}^{(k)}_{\mathrm{rel}} = 0$.
In this case, $I_0 = I_\mathrm{Landauer}$ reduces to the conventional Landauer form for transport through a static barrier with transmission function $\mathcal{T}^{(0)}_{\mathrm{avg}} \equiv \mathcal{T}$ \cite{cuevas2nd}.
    
Additional degrees of freedom, e.g., spin or internal lattice sites,  are considered through the trace $\mathrm{Tr}(\cdot)$.
The matrices $\Gamma_a$ define how both terminals couple to these degrees of freedom.
In the wide-band limit, they are energy-independent and introduce a generally inhomogeneous spectral broadening of the energy levels of the center \cite{cuevas2nd}. 
Finally, the dressed retarded Green's function elements $\mathcal{G}_{k}$ are determined from the Dyson equation $\mathcal{G}_{k}(E) = \delta_{k0} g_0(E) + g_k(E) \sum_{l \in\mathbb{Z}} W_{k-l} \mathcal{G}_{l}(E)$ \cite{supmat,Aoki2014,Liu2017,Mosallanejad2024b}, with the bare retarded Green's function $g_k(E) = [ (E-k\hbar\Omega)\mathbbm{1} - H_C + i (\Gamma_L + \Gamma_R)/2]^{-1}$ of the undriven center, already including the coupling with the contacts.

\sect{Measurement proposal}As sketched in Fig.~\ref{fig:fig1}(b), the QGT transition elements $q_{\alpha\beta,pn}$ in Eq.~\eqref{eq:qgt} naturally arise from conductance measurements in linear response.
Assuming that the voltage $V$ across the junction is small around a fixed chemical potential $\mu$ in the terminals, $\mu_{L,R} = \mu \pm eV/2$, the dc current in Eq.~\eqref{eq:averageCurrent} becomes $I_0(\mu) = I_\mathrm{pump}(\mu) + G(\mu) V + \mathcal{O}(V^2)$.
At sufficiently low temperatures, the linear conductance $G(\mu) = (e^2/h) \sum_{k \in \mathbb{Z}} \mathcal{T}^{(k)}_{\mathrm{avg}}(\mu)$ will only depend on the average chemical potential $\mu$ (see End Matter, Appendix A).
Corrections for higher temperatures can be taken into account with a Sommerfeld expansion \cite{Benenti2017}.

We define $G_n \equiv G(\varepsilon_n)$ for $\mu = \varepsilon_n$ as the conductance through an energy level in the center. 
Weakly driving the the center parameters, $A \ll 1$, the dc current can be written as $I_0(\varepsilon_n) \approx I_\mathrm{pump}(\varepsilon_n) + (G_n|_{A=0} + \delta G_n)  V$, where both $I_\mathrm{pump}(\varepsilon_n) \propto A^2$ and $\delta G_n \propto A^2$. 
The conductance correction $\delta G_n$ originating from the drive can now be isolated by subtracting two reference measurements that determine $I_\mathrm{pump}(\varepsilon_n)$ and $G_n|_{A=0}$ at $V = 0$ and $A = 0$, respectively. 

As derived in the SM \cite{supmat}, driving resonantly with $\hbar \Omega = |\varepsilon_n - \varepsilon_p|$ $(p \neq n)$ leads to a dominant contribution in $\delta G_n$ that is proportional to the QGT transition element $q_{\alpha\beta,pn}$.
Weakly modulating one parameter $\lambda_\alpha$ leads to $\delta G_n \approx (e^2/h) A^2 P_{pn}  \, q_{\alpha\alpha,pn}$, with
\begin{align}
\label{eq:parameterDependentPrefactors}
    P_{pn} 
    &= 
    - \frac{ (\varepsilon_p - \varepsilon_n)^2 }{  4 \Gamma_p \Gamma_n } 
    (1  + \delta_p \delta_n - 2 \delta_n^2) .
\end{align}  
Here, we define the average broadening $\Gamma_n = (\Gamma_{L,n} + \Gamma_{R,n})/2$ and the relative difference $\delta_n = (\Gamma_{L,n} - \Gamma_{R,n})/(2\Gamma_{n})$.
In general, $P_{pn}$ depends on $\lambda$ and takes into account the details of how the contacts couple to the eigenstates of the undriven center.

The simplified formula in Eq.~\eqref{eq:parameterDependentPrefactors} includes the dominant terms for $k = -1,0,1$ in the tunneling regime, in which it is sufficient to use a secular approximation and restrict the calculation to the diagonal elements $\Gamma_{a,n} \equiv \braket{n|\Gamma_a|n}$.
For stronger couplings, different transport channels get mixed and all elements $\braket{m|\Gamma_a|n}$ have to be taken into account in the trace formula for $\mathcal{T}_{ab}^{(k)}$. 
From the experimental side, all $\Gamma_{a,n}$ should be determined from the peak width of $G(\mu)|_{A=0}$ around $\mu \approx \varepsilon_n$, which, in the tunneling limit, can be modeled by a sum of Lorentzians with fitting parameters $\Gamma_{a,n}$ (see End Matter, Appendix B).

Driving two parameters $(\lambda_\alpha,\lambda_\beta)$ $(\alpha\neq \beta)$ simultaneously, we get the change in conductance $\delta G_{n}^{(\gamma)}$ that depends on the choice of the relative phase $\gamma$ of the drive.
While the difference between two measurements at $\gamma = 0$ and $\gamma = \pi$ (orthogonal linear drives) yields $\delta G_n^{(0)} - \delta G_n^{(\pi)} \approx  4 (e^2/h) A^2 P_{pn} \mathrm{Re}(q_{\alpha\beta,pn})$, the difference between two measurements at $\gamma = \pm\pi/2$ (orthogonal circular drives) yields $\delta G_n^{(-\pi/2)} - \delta G_n^{(+\pi/2)} \approx 4  (e^2/h) A^2 P_{pn} \mathrm{sgn}(\varepsilon_p - \varepsilon_n)   \mathrm{Im}(q_{\alpha\beta,pn})$. 

\begin{figure}
    \centering
    \includegraphics{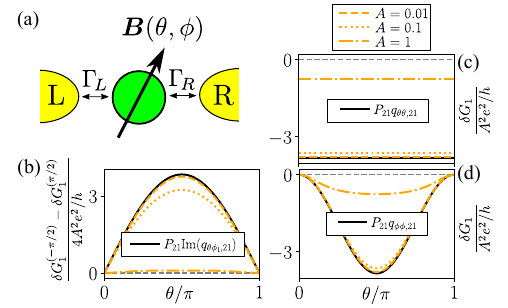}
    \caption{(a) Two-level system modeled as a spin in a magnetic field $\boldsymbol{B}(\theta,\phi)$. (b--d) Conductance correction $\delta G_1$ through the energy level $\varepsilon_1 = - B$. Solid lines represent the nonzero elements of the QGT $q_{\alpha\beta,21}$ in Eq.~\eqref{eq:QGTtwolevelsystem}, other lines are numerical results for certain driving amplitudes $A$ and different one-and two-parameter drives. Parameters: $\phi = 0$, $\Gamma = B/4$, $\delta = 1/5$, $\mu = -B$, $\hbar \Omega = 2B$.}
    \label{fig:fig2}
\end{figure}

\sect{Example 1: Homogeneously coupled two-level system}To give a first simple example, we consider a generic two-level system ($N = 2$), $H_C = \sum_{i=1}^3 B_i \sigma_i$, where $\sigma_i$ are Pauli matrices and $B_i$ are functions  that map to a sphere of radius $B = \sqrt{B_1^2+B_2^2+B_3^2} > 0$. 
The parameters of the model are the spherical angles $(\lambda_1,\lambda_2) = (\theta,\phi) \in [0,\pi] \times [0,2\pi)$ (see End Matter, Appendix C).
As sketched in Fig.~\ref{fig:fig2}(a), an experimental setup could be a single noninteracting quantum dot  with an external magnetic field of strength $B$, where the angles are controlled relative to a global spin quantization axis.
The energies of the system are $\varepsilon_{1,2} = \mp B$ and the QGT transition elements read \cite{Kolodrubetz2017}
\begin{align}
    \label{eq:QGTtwolevelsystem}
    (q_{\alpha\beta,21}) = 
    \frac{1}{4} 
    \begin{pmatrix}
        1 & - i \sin (\theta ) \\
        i \sin (\theta ) & \sin ^2(\theta )  \\
    \end{pmatrix} 
     .
\end{align}
Since $q_{\alpha\beta,1} \equiv q_{\alpha\beta,21} = q_{\alpha\beta,12}^* \equiv q_{\alpha\beta,2}^*$, we  focus on state $\ket{1}$ and determine the conductance correction $\delta G_1$ through this energy level, while driving on resonance $\hbar \Omega = 2 B$. 
For simplicity, we assume that the terminals couple equally to spin up and down, i.e., $\Gamma_a \equiv \Gamma_a \mathbbm{1}$ are scalars.
Then, $\mathcal{T}_\mathrm{avg}^{(k)} \equiv  \Gamma_R \Gamma_L  \mathrm{Tr}(  \mathcal{G}_{k} \mathcal{G}_{k}^\dag)$ and $\mathcal{T}_\mathrm{rel}^{(k)} = 0$, since $ \mathcal{T}_{RL}^{(k)} = \mathcal{T}_{LR}^{(k)} $ for all modes $k$.
Hence, there is no pump current, $I_\mathrm{pump} = 0$,
and Eq.~\eqref{eq:parameterDependentPrefactors} simplifies to $P_{21} = P_{12} = - B^2 (1-\delta^2) / \Gamma^2$, since $\Gamma_n \equiv \Gamma$ and $\delta_n \equiv \delta$.

We numerically calculate $\mathcal{T}_\mathrm{avg}^{(k)}$ for different driving amplitudes $A$ by solving the Dyson equation for the retarded elements $\mathcal{G}_k$ including terms $|k| \leq 6$.
The resulting change in conductance $\delta G_1$ and their relation with the QGT elements in Eq.~\eqref{eq:QGTtwolevelsystem} are shown in Figs.~\ref{fig:fig2}(b)--\ref{fig:fig2}(d).

We see that the numerical results are well described by the approximation discussed around Eq.~\eqref{eq:parameterDependentPrefactors} for weakly modulated parameters ($A = 0.01$, dashed lines), and allow to relate the conductance correction $\delta G_1$ with all  QGT elements $q_{\alpha\beta,21}$ in the tunneling regime. 
An increase in the coupling strength $A$ leads to a violation of the weak-modulation regime, and requires the inclusion of higher-order corrections beyond $\propto A^2$.
For $A = 1$ (dash-dotted lines), the result cannot be related with the elements of the QGT in a simple way. 

Finally, we determine the Chern number $C$ of state $\ket{1}$, which can be computed from either the real or imaginary part of the measured QGT $q_1 \equiv q_{21}$ \cite{Palumbo2018,Palumbo2019,Yu2019,Ozawa2021,Mera2021}.
Using the numerical results for a driving amplitude of $A = 0.01$, we obtain 
$C_{\mathrm{Re}(q_1)} \approx 0.986$ and $C_{\mathrm{Im}(q_1)} \approx 0.978$, which both agree very well with the analytical result $C = 1$ (see End Matter, Appendix C). 

\sect{Example 2: Inhomogeneously coupled three-level system}Since for the two-level system there is only one resonance frequency and the QGT of the two states are complex conjugates of each other, we apply the method to a three-level system ($N=3$) to discuss the more general aspects of our proposal.
To be able to compare the numerical results with an exact analytical QGT, we use a minimal three-level model that exhibits a so-called tensor monopole \cite{Palumbo2018,Palumbo2019,Weisbrich2021b,Tan2021,Chen2022,jankowski2025}.

\begin{figure}
	\centering
	\includegraphics{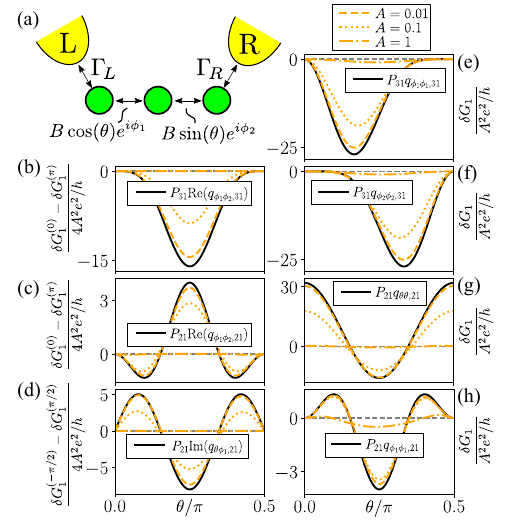}
	\caption{(a) Three-level system with couplings to the terminals. Green circles represent lattice sites at zero onsite potential. (b--h) Conductance correction $\delta G_1$ through the energy level $\varepsilon_1 = - B$. Black solid lines represent the elements of the QGT in Eq.~\eqref{eq:qgt3levelsystem}, orange lines are numerical results for different driving amplitudes $A$ and different one-and two-parameter drives on resonance: (b,e,f) $\hbar \Omega = 2B$, (c,d,g,h) $\hbar \Omega = B$. Common parameters: $\phi_1 = \phi_2 = 0$, $\Gamma = B/4$, $\delta = 0$, $\mu = -B$.}
	\label{fig:fig3}
\end{figure}

Translated to a transport setup, we consider a chain of three noninteracting quantum dots with zero onsite energies, as sketched in Fig.~\ref{fig:fig3}(a). 
Since only the two outer sites are connected with terminals, this represents an inhomogeneous coupling situation, as is typical for electronic transport setups. 
The interdot hoppings $B$ shall be modulated by $\cos(\theta)$ and $\sin(\theta)$, and the electrons acquire hopping phases $\phi_1$ and $\phi_2$.
The energies are $\varepsilon_{1,3} = \mp B$ and $\varepsilon_2 = 0$, where $B = \sqrt{B_1^2 + B_2^2 + B_3^2 + B_4^2}$, and the parameterization reads 
$B_1 = B \cos(\theta) \cos(\phi_1)$, $B_2 = B \cos(\theta) \sin(\phi_1)$, $B_3 = B \sin(\theta) \cos(\phi_2)$ $B_4 = B \sin(\theta) \sin(\phi_2)$, where $B > 0$, and $(\lambda_1,\lambda_2,\lambda_3) = (\theta,\phi_1,\phi_2) \in [0,\pi/2] \times [0,2\pi)^2$ are the parameters (see End Matter, Appendix D). 
The QGT transition elements are \cite{Chen2022}
\begin{align}
\label{eq:qgt3levelsystem}
\begin{split}
(q_{\alpha\beta,21}) 
=
(q_{\alpha\beta,23}) 
&=
\begin{pmatrix}
 1/2  & i c s & -i c s \\
 -i c s & 2 c^2 s^2 & -2 c^2 s^2 \\
 i c s & -2 c^2 s^2 & 2 c^2 s^2 \\
\end{pmatrix} ,
 \\
(q_{\alpha\beta,31})
&= 
\begin{pmatrix}
 0 & 0 & 0 \\
 0 & c^4 & c^2 s^2 \\
 0 & c^2 s^2 & s^4 \\
\end{pmatrix} ,
\end{split}
\end{align} 
with $c(\theta) = \cos (\theta )/\sqrt{2}$ and $s(\theta) = \sin (\theta )/\sqrt{2}$.

For simplicity, we consider both coupling strengths equal to $\Gamma$.
In this case, Eq.~\eqref{eq:parameterDependentPrefactors} simplifies to $P_{12} = P_{21} = P_{23} = P_{32} = B^2 [ 1+3 \cos (4 \theta )] / \Gamma^2$ and $	P_{13} = P_{31} = - 8 B^2  [ 1-\cos (4 \theta )] / \Gamma ^2$.
For completeness, the general expressions for $P_{pn}$ for unequal coupling strengths, which does not change our results qualitatively, is provided in the End Matter, Appendix D.

Again, we numerically calculate $\mathcal{T}_\mathrm{avg}^{(k)}$ for arbitrary driving amplitudes $A$ by solving the Dyson equation for the retarded elements $\mathcal{G}_k$ up to order $|k| \leq 6$. 
The resulting conductance correction $\delta G_1$ at $\mu = \varepsilon_1$ at the two resonance frequencies $\hbar \Omega = B$ and $\hbar \Omega = 2B$, and their relations with the analytical QGT, are shown in Figs.~\ref{fig:fig3}(b)--\ref{fig:fig3}(h).

The results match the expected analytical results well for small driving amplitude ($A = 0.01$, orange dashed lines), and the agreement gets worse for stronger driving in complete analogy with the two-level system in Fig.~\ref{fig:fig2}
In addition, the inhomogeneous broadening $\Gamma_n$ along the chain now induces an additional parameter dependence in $P_{pn}$.
For instance, since $q_{\theta\theta,21} = 1/2$ is constant, Fig.~\ref{fig:fig3}(g) essentially shows how $P_{21}$ depends on $\theta$. 
In general transport setups, $P_{pn}$ can have roots for which $P_{pn}(\lambda) = 0$, which in turn does not allow directly identifying the QGT transition elements at these isolated parameter points. 
However, it is still possible to deduce these elements for all values away from the roots of $P_{pn}$. 

Finally, we determine the Dixmier-Douady invariant $\mathcal{DD}$ of state $\ket{1}$, which can also be determined from either the real or imaginary part of the measured QGT $q_1 \equiv q_{21} + q_{31}$ \cite{Palumbo2018,Palumbo2019,Chen2022}.
Using the numerical results for $A = 0.01$, we obtain $\mathcal{DD}_{\mathrm{Re}(q_1)} \approx 0.930$ and $\mathcal{DD}_{\mathrm{Im}(q_1)} \approx 0.996$.
While $\mathcal{DD}_{\mathrm{Im}(q_1)}$ matches the analytical result $\mathcal{DD} = 1$ quite well, $\mathcal{DD}_{\mathrm{Re}(q_1)}$ shows a larger deviation, which originates from errors caused by the isolated roots of $P_{pn}$.
These errors mostly cancel for $\mathcal{DD}_{\mathrm{Im}(q_1)}$ under integration in the sense of a principal value (see End Matter, Appendix D).

\sect{Conclusion}We have presented and discussed a general protocol for reading out the full quantum geometry of the quantum states of a transport device.
By combining the ideas of parametric pumping with specifically chosen resonant periodic drives of one or two parameters of the center region, we have shown how the change in conductance across the device is dominated by the transition elements of the QGT defined in Eq.~\eqref{eq:qgt} [cf.~Fig.~\ref{fig:fig1}].
Most importantly, we have discussed how quantum geometry can be isolated through various conductance measurements in general transport setups, such as tunnel or molecular junctions, which typically involve inhomogeneous couplings between the device and the terminals. 
In the tunneling regime, these effects can be approximated with a generally parameter-dependent factor $P_{pn}$ [Eq.~\eqref{eq:parameterDependentPrefactors}], which is characteristic to the investigated device. 
We have illustrated these general results with specifically engineered two-level and three-level systems, which allowed us to compare the numerical calculation of the conductance with the exact analytical expressions of the QGT [cf.~Figs.~\ref{fig:fig2} and \ref{fig:fig3}]. 
In general, the parameter dependence of $P_{pn}$ is reduced or even eliminated when the terminals couple to the device homogeneously [cf.~Fig.~\ref{fig:fig2}].

By providing a simple and clean strategy to measure the QGT in transport systems, our theory paves the way for more sophisticated applications of quantum geometry in quantum electronic devices.
We believe that this work will inspire both theoretical and experimental efforts to extend and test these concepts across a wide range of transport setups, such as novel hybrid superconductor-semiconductor devices \cite{Prada2020} or geometric effects in molecular junctions \cite{steinmetz2025}.

\begin{acknowledgments}
    We thank Nathan Goldman,  Gonzalo Usaj, Lucila Peralta Gavensky, and Sigmund Kohler for stimulating discussions. 
    M.B. acknowledges funding from the Horizon Europe Framework Program of the European Commission through the European Innovation Council Pathfinder Grant No.~101115315 (QuKiT).
\end{acknowledgments}

\bibliography{refs.bib}

\clearpage

%
%
\section*{End Matter}
\sect{Appendix A: Linear conductance at low temperature}
Using $\mu_{L,R} = \mu \pm eV/2$ for the chemical potentials in the contacts and assuming that the applied voltage $V$ around $\mu$ is small, we expand the Fermi functions
\begin{align}
	f_{L,R}(E) 
	&=
	f(E) 
	\mp 
	\frac{\partial f}{\partial E} \, \frac{eV}{2}
	+
	\mathcal{O}(V^2) ,
\end{align}
with $f(E) = [1 + \exp(\beta(E-\mu)) ]^{-1}$. 
The two contributions to the dc current $I_0$ in Eq.~\eqref{eq:averageCurrent} of the main text become 
\begin{subequations}
\begin{align}
	I_\mathrm{Landauer}
	&= 
	\frac{e^2}{h}  
	\int_\mathbb{R} dE  
	\sum_{k \in \mathbb{Z}}    
	\mathcal{T}_\mathrm{avg}^{(k)}(E)
	\left( 
	- 
	\frac{\partial f}{\partial E} \right) V  
	+
	\mathcal{O}(V^3) ,
\\ 
	I_\mathrm{pump}
	&=
	\frac{2 e}{h}  
	\int_\mathbb{R} dE  
	\sum_{k \in \mathbb{Z}}    
	\mathcal{T}_\mathrm{rel}^{(k)}(E) 
	f(E) + \mathcal{O}(V^2) 
	.
\end{align}  
\end{subequations}
In the low-temperature limit $\beta \to \infty$, we have $f(E) \to \theta(\mu - E)$ and $( - \partial f / \partial E) \to \delta(E-\mu)$, which results in
\begin{subequations}
\begin{align} 
	I_\mathrm{Landauer}(\mu)
	&= 
	G(\mu)  V  
	+
	\mathcal{O}(V^3) , 
	\\
	I_\mathrm{pump}(\mu)
	&=
	\frac{2 e}{h}  
	\int_{-\infty}^\mu dE  
	\sum_{k \in \mathbb{Z}}    
	\mathcal{T}_\mathrm{rel}^{(k)}(E) 
	+ \mathcal{O}(V^2)
	,
\end{align}
\end{subequations}
with the linear conductance 
\begin{align}
	\label{eq:linearConductance}
	G(\mu) = \frac{e^2}{h}   
	\sum_{k \in \mathbb{Z}}    
	\mathcal{T}_\mathrm{avg}^{(k)}(\mu) .
\end{align}

\sect{Appendix B: Conductance in the static tunneling limit}In the static case without the drive, i.e., $A = 0$ and $W_k = 0$, we have $I_\mathrm{pump} = 0$, while $\mathcal{T}_\mathrm{avg}^{(k)} = 0$ for $k \neq 0$ and $\mathcal{T}^{(0)}_{\mathrm{avg}} = \mathrm{Tr}( \Gamma_R g_{0} \Gamma_L g_{0}^\dag)$.
In the tunneling limit with weakly coupled terminals, the bare retarded Green's function of the center, including the coupling with the leads, can be well approximated by the diagonal Lehmann representation
\begin{align} 
    \label{eq:apprixLehmann}
	g_k(E) \approx \sum_n \frac{ \ket{n}\bra{n} }{
		E- k \hbar \Omega  - \varepsilon_n + i \Gamma_n} ,
\end{align}
where $\ket{n}$ and $\varepsilon_n$ are the eigenstates and eigenenergies of the center Hamiltonian $H_C(\lambda)$, while $\Gamma_n = (\Gamma_{L,n} + \Gamma_{R,n})/2$ and $\Gamma_{a,n} = \braket{n|\Gamma_a|n}$ are the broadenings of the states (secular approximation). 
Note that Eq.~\eqref{eq:apprixLehmann} becomes exact, even beyond the tunneling limit, if $\braket{m|\Gamma_a|n} = 0$ for $m\neq n$, which, however, is typically not the case for general transport setups. 

Therefore, in the static tunneling (and secular-approximated) limit, the conductance in Eq.~\eqref{eq:linearConductance} simplifies to a sum of Lorentzians, 
\begin{align}
	G(\mu)|_{A=0} \approx \frac{e^2}{h}      
    \sum_n  \frac{ \Gamma_{R,n} \Gamma_{L,n}  }{(\mu  - \varepsilon_n)^2 + \Gamma_n^2 } .
\end{align}
This serves as a good approximation for chemical potentials in the region of the transmission peaks, $\mu \approx \varepsilon_n$, and can be used for fitting the measured conductance to obtain experimental values for $\Gamma_{a,n}$.

\sect{Appendix C: Two-level system with uniform coupling}Example 1 in the main text is described by the Hamiltonian
\begin{align}
	\label{eq:HamiltonianTwoLevelSystem}
	H_C = \begin{pmatrix}
		B \cos(\theta) & B \sin(\theta) e^{-i\phi} \\
		B \sin(\theta) e^{i\phi} & -B \cos(\theta) \\
	\end{pmatrix} , 
\end{align}
with eigenenergies $\varepsilon_{1,2} = \mp B$ and eigenstates
\begin{align}
	\ket{1} 
	&= \begin{pmatrix}
		e^{-i \phi } \sin(\theta/2)
		\\
		-\cos( \theta / 2)
	\end{pmatrix} 
	, 
	\ket{2} 
	= \begin{pmatrix}
		e^{-i \phi } \cos(\theta/2)
		\\
		\sin( \theta / 2)
	\end{pmatrix} .
\end{align}
The states are used to determine the QGT transition elements in Eq.~\eqref{eq:QGTtwolevelsystem} according to Eq.~\eqref{eq:qgt}.
The coupling matrices $\Gamma_{L,R} \equiv \Gamma (1\pm \delta) \mathbbm{1}$ describe homogeneous couplings with the terminals. 
The level broadenings $\Gamma_{a,n} = \braket{n|\Gamma_a|n} \equiv \Gamma_a$ are the same for both energy levels, which reduces $P_{pn}$ in Eq.~\eqref{eq:parameterDependentPrefactors} directly to 
the ones presented in the main text.

The Berry curvature $F_{\theta\phi,1}$ of the low-energy state $\ket{1}$ is directly obtained from the QGT in Eq.~\eqref{eq:QGTtwolevelsystem} in the main text as $F_{\theta\phi,1} 
= - 2 \, \mathrm{Im}(q_{\theta\phi,1}) = \sin (\theta )/2$, which determines the topological Chern number 
\begin{align}
    C = \frac{1}{2\pi} \int_0^\pi d\theta \int_{0}^{2\pi} d\phi \, F_{\theta\phi,1} = 1.
\end{align}
As pointed out in Ref.~\cite{Ozawa2021}, the Chern number of the two-level system is directly related with the so-called quantum volume, calculated from the determinant of the Fubini-Study metric $\mathrm{Re}(q_{\alpha\beta,1})$.
In other words, we can also determine the Berry curvature as \cite{Palumbo2018,Ozawa2021}
\begin{align}
\label{eq:appendixBerryCurvatureReQGT}
    F_{\theta\phi,1} = 2 \sqrt{ \mathrm{det}( \mathrm{Re}(q_{1}) ) }.
\end{align}

\begin{figure*}
	\centering
    \includegraphics{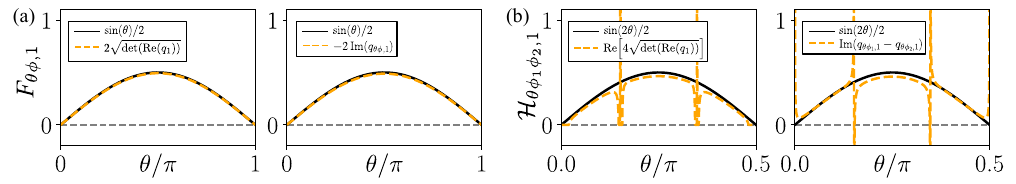}
	\caption{Numerical results (orange dashed lines) for $A=0.01$ for (a) the Berry curvature $F_{\theta\phi,1}$ (two-level system) and (b) the tensor Berry curvature $\mathcal{H}_{\theta\phi_1\phi_2,1}$ (three-level system) for the lowest-energy state $\ket{1}$.
    The analytical results are shown as black solid lines. 
    In each panel, the left plot shows the result obtained from the determinant of the Fubini-Study metric (real part of the QGT), while the right plot shows the result from the Berry curvatures (imaginary part of the QGT).
    Parameters: In panels (a) and (b), we use the same parameters as in Figs.~\ref{fig:fig2} and \ref{fig:fig3} of the main text, respectively.
    }
	\label{fig:fig4}
\end{figure*}

As explained in the main text, engineered linear and circular resonant drives of the parameters at $\hbar \Omega = 2B$ allow to reconstruct the QGT elements $q_{\alpha\beta,21} \equiv q_{\alpha\beta,1}$ from a change in conductance $\delta G_1$ through state $\ket{1}$.
The two numerical results for the Berry curvature obtained from the real and imaginary parts of the QGT for a driving amplitude $A=0.01$ are shown in Fig.~\ref{fig:fig4}(a).
We see that both results agree very well with the analytical form of the Berry curvature.
The Chern number calculated from the real part of the QGT yields 
$C_{\mathrm{Re}(q_1)} \approx 0.986$, while the result from the imaginary part results in $C_{\mathrm{Im}(q_1)} \approx 0.978$.

\sect{Appendix D: Three-level system with nonuniform coupling}Example 2 in the main text is described by the Hamiltonian
\begin{align}
	\label{eq:HamiltonianThreeLevelSystem}
	H_C = \begin{pmatrix}
		0 & B \cos(\theta) e^{-i\phi_1} & 0  \\
		B \cos(\theta) e^{i\phi_1} & 0 & B \sin(\theta) e^{i\phi_2} \\
		0 & B \sin(\theta) e^{-i\phi_2} & 0 
	\end{pmatrix} ,
\end{align}
with eigenenergies $\varepsilon_{1,3} = \mp B$, $\varepsilon_2 = 0$, and eigenstates
\begin{align}
    \label{eq:eigenstates3BandModel}
	\ket{1,3} 
	&= 
	\frac{1}{\sqrt{2}}
	\begin{pmatrix}
	e^{-i \phi_1} \cos (\theta )
	\\
	\mp 1
	\\
	e^{-i \phi_2} \sin (\theta ) 
	\end{pmatrix}, 
	\ket{2} 
	= 
	\begin{pmatrix}
		-e^{i \phi_2} \sin (\theta )
		\\
		0
		\\
		e^{i \phi_1} \cos (\theta )
	\end{pmatrix}
	 .
\end{align}
The couplings are inhomogeneous,
\begin{align} 
	\Gamma_{L} \equiv \begin{pmatrix}
		\Gamma (1+\delta) & 0 & 0 \\
		0 & 0 & 0 \\
		0 & 0 & 0 
	\end{pmatrix} , 
	\Gamma_{R} \equiv \begin{pmatrix}
		0 & 0 & 0 \\
		0 & 0 & 0 \\
		0 & 0 & \Gamma (1-\delta) 
	\end{pmatrix} ,
\end{align}
and determine the broadenings 
$\Gamma_{a,n} \equiv \braket{n|\Gamma_a|n}$
that enter Eq.~\eqref{eq:parameterDependentPrefactors} to determine the prefactors $P_{pn}$. 
In detail, they read
\begin{subequations}
\begin{align}
	\Gamma_{L,1} = \Gamma_{L,3} &= \Gamma  (1 + \delta) \cos ^2(\theta ) / 2 , 
    \\
	\Gamma_{L,2} &= \Gamma  (1 + \delta) \sin ^2(\theta ), \\
	\Gamma_{R,1} = \Gamma_{R,3} &=  \Gamma  (1-\delta) \sin ^2(\theta ) / 2 , 
    \\
	\Gamma_{R,2} &= \Gamma  (1 - \delta) \cos ^2(\theta ) .
\end{align} 
\end{subequations}
This results in $P_{12} = P_{32} = P(\delta)$, $P_{21} = P_{23} = P(-\delta)$, and $P_{13} = P_{31}$, with 
\begin{subequations}
\begin{align}
	P(\delta) &= \frac{B^2 (1-\delta ^2)}{2\Gamma^2}   \,   \frac{2+6 \cos (4 \theta ) -9 \delta  \cos (2 \theta )+\delta  \cos (6 \theta )}{ ( 1 - \delta  \cos (2 \theta ))^3 ( 1 +  \delta  \cos (2 \theta ) )^2} ,
	\\
	P_{13} &= - \frac{8 B^2}{\Gamma^2} (1-\delta ^2) \,  \frac{ 1-\cos (4 \theta ) }{(1 + \delta  \cos (2 \theta ))^4} .
\end{align}
\end{subequations}
The case $\delta = 0$ is presented and discussed in the main text. A finite asymmetry $\delta \in (-1,1)$ shifts the positions of the roots of $P_{pn}$ in parameter space, which does not change our findings qualitatively.

Following Refs.~\cite{Palumbo2018,Palumbo2019}, we generally calculate the tensor Berry curvature 
\begin{align}
    \mathcal{H}_{\theta\phi_1\phi_2,1}
    =
    \partial_\theta B_{\phi_1\phi_2,1}
    +
    \partial_{\phi_1} B_{\phi_2\theta,1}
    +
    \partial_{\phi_2} B_{\theta\phi_1,1}
\end{align}
of the lowest-energy state $\ket{1}$ from its tensor Berry connection 
\begin{align}
    B_{\alpha\beta,1}
    =
    \frac{i}{3} \sum_{a,b,c=1}^3 
    \varepsilon_{abc} \varphi_a (\partial_\alpha \varphi_b) (\partial_\beta \varphi_c),
\end{align}
with the Levi-Civita symbol $\varepsilon_{abc}$.
The three scalar fields $\varphi_a$ are defined as $\varphi_1 = - i \ln(v_2)$, $\varphi_2 = v_1^*$, and $\varphi_3 = v_1$, with the nontrivial components of state $\ket{1}$ in Eq.~\eqref{eq:eigenstates3BandModel},
$v_1 = e^{-i \phi_1} \cos (\theta ) / \sqrt{2}$ and $v_2 = e^{-i \phi_2} \sin (\theta ) / \sqrt{2}$.
This results in $\mathcal{H}_{\theta\phi_1\phi_2,1} = \sin (2 \theta ) / 2$ and the topological Dixmier-Douady (DD) invariant 
\begin{align}
    \mathcal{DD} = \frac{1}{2\pi^2} \int_0^{\pi/2} d\theta \int_{0}^{2\pi} d\phi_1  \int_{0}^{2\pi} d\phi_2 \,  \mathcal{H}_{\theta\phi_1\phi_2,1} = 1 .
\end{align}

As pointed out in Refs.~\cite{Palumbo2018,Chen2022}, the DD invariant can also be determined from the QGT for the three-level system that we use as an example. 
By measuring the QGT elements $q_{\alpha\beta,21}$ and $q_{\alpha\beta,31}$ from a change in conductance $\delta G_1$ through state $\ket{1}$ under two resonant drives $\hbar \Omega = B$ and $\hbar \Omega = 2B$, respectively, as explained in the main text, the QGT $q_{\alpha\beta,1}$ of state $\ket{1}$ follows as $q_{\alpha\beta,1} = q_{\alpha\beta,21} + q_{\alpha\beta,31}$ [cf.~Eq.~\eqref{eq:qgt} in the main text]. 
Then, we can also use $\mathcal{H}_{\theta\phi_1\phi_2,1}
   =
   4 \sqrt{ \mathrm{det}( \mathrm{Re}(q_{1}) ) }$
and $ 
   \mathcal{H}_{\theta\phi_1\phi_2,1}
   =
   \mathrm{Im}(q_{\theta\phi_1,1} - q_{\theta\phi_2,1})
$, which allow for direct experimental measurement \cite{Chen2022}. 

The two numerical results obtained from the real and imaginary parts of the QGT for a driving amplitude $A=0.01$ are shown in Fig.~\ref{fig:fig4}(b).
We see that both results agree quite well with the analytical results away from isolated points in parameter space for which $P_{pn} = 0$. Close to these points, we encounter divergencies that lead to strong deviations and partially to the violation of basic properties of the QGT, e.g., regions where $\mathrm{det}( \mathrm{Re}(q_{1}) ) < 0$ such that $\sqrt{\mathrm{det}( \mathrm{Re}(q_{1}) )}$ is imaginary.
Integrating the parameter regions for which $\mathrm{det}( \mathrm{Re}(q_{1}) ) \geq 0$ results in $\mathcal{DD}_{\mathrm{Re}(q_1)} \approx 0.930$, while the results from the imaginary part of the QGT results in $\mathcal{DD}_{\mathrm{Im}(q_1)} \approx 0.996$.
In particular, the result from the imaginary part agrees very well with the expected analytical result, since the integration over the divergencies cancel each other in the sense of a principal value
[$P_{21}$ crosses zero linearly, cf.~Fig.~\ref{fig:fig3}(g) in the main text].
Such a cancellation is not possible for the calculation from the determinant formula of the real part, resulting in a larger deviation.

\clearpage

%
%
\begin{widetext}
\renewcommand{\theequation}{S\arabic{equation}}
\renewcommand{\theHequation}{S\arabic{equation}}
\setcounter{equation}{0}
{\centering
\textbf{
\large
--- Supplemental Material --- 
\\\vspace{0.2cm}
\manuscriptTitle
}
\\
\vspace{0.4cm}
\normalsize Raffael~L.~Klees and M{\'o}nica~Benito
\\
{\small \textit{\UniA} 
\\
{(Dated: \today)} }
\\
}

\section{Current across a driven center}
As introduced in the main text, the general transport device is defined by a Hamiltonian of the form $H = H_L + H_R + H_C(\lambda) + H_{LC} + H_{RC}$.
The electronic current $I_a$ in terminal $a = L,R$ can be defined as the change of particle number in the respective terminal via the Heisenberg equation of motion \cite{cuevas2nd}, i.e.,
\begin{align}
	I_a(t) = e \Braket{ \frac{d }{dt} N_a(t) } = \frac{ie}{\hbar} \Braket{ [H(t), N_a(t)] } ,
	\label{eq:chargeCurrentDefinition}
\end{align}
for which operators are defined in the Heisenberg picture. 
Here, $e$ and $\hbar$ are the electron charge and the reduced Planck constant, respectively. 

We assume that the terminals are normal metals and described as free noninteracting electrons.
The couplings between the center and the terminals are assumed to be spin-independent and defined by the Hamiltonian
\begin{align}
	H_{aC} = \sum_{\sigma\sigma'} (c^\dag_{a\sigma} v_{aC,\sigma\sigma'} c_{C\sigma'} +  c_{C\sigma}^\dag v_{Ca,\sigma\sigma'} c_{a\sigma'}),
\end{align}
where $c_{C\sigma}^{(\dag)}$ are annihilation (creation) operators in the center, $c_{a\sigma}^{(\dag)}$ are annihilation (creation) operators in terminal $a = L,R$, and $v_{aC,\sigma\sigma'}$ and $v_{Ca,\sigma\sigma'}$ are the couplings between the two subsystems. 
Note that the index $\sigma$ is a short-hand notation of additional degrees of freedom, such as, for instance, lattice sites, orbitals, spin, and so on.

Only the coupling parts $H_{aC}$ of the Hamiltonian $H$ lead to a nonzero contribution when calculating the commutator in Eq.~\eqref{eq:chargeCurrentDefinition}.
The result is  
\begin{align}
	I_a(t) 
	&=  e \,  \mathrm{Tr}\Bigl(   
	V_{Ca} G^<_{aC}(t,t) 
	-  
	V_{aC} G^<_{Ca}(t,t)
	\Bigr)  , 
	\label{eq:currentStartingPoint}
\end{align} 
where we pack the additional degrees of freedom $\sigma$ into matrices, e.g., $V_{aC} = (v_{aC,\sigma\sigma'})_{\sigma\sigma'}$.
The trace and matrix product takes into account the summation over $\sigma$. 
In passing by, we defined the lesser Green's function 
\begin{align}
	G_{ab,\sigma\sigma'}^<(t,t') = \frac{i}{\hbar} \braket{c^\dag_{b\sigma'}(t') c_{a\sigma}(t)}.
\end{align}
In the following, we will use the nonequilibrium Keldysh formalism to determine the full dressed lesser Green's function.
In general, all dressed Green's functions $G$ will be calculated from the bare Green's functions $g$ of the uncoupled subsystems, $H_L$, $H_R$, and $H_C$, by means of Dyson equations.
To complete the set of Green's functions that we will need in the rest of this derivation, we also introduce the greater Green's function
\begin{align}
	G^>_{ab,\sigma\sigma'}(t,t') = - \frac{i}{\hbar} \braket{c_{a\sigma}(t) c^\dag_{b\sigma'}(t')},
\end{align}
as well as the retarded and advanced Green's functions 
\begin{subequations}
	\label{eq:retardedAdvancedDefinitions}
\begin{align}
	G^\mathrm{ret}_{ab,\sigma\sigma'}(t,t') &= - \frac{i}{\hbar} \theta(t-t') \braket{ \{ c_{a\sigma}(t) ,c^\dag_{b\sigma'}(t') \} }, 
	\\
	G^\mathrm{adv}_{ab,\sigma\sigma'}(t,t') &= \frac{i}{\hbar} \theta(t'-t) \braket{ \{ c_{a\sigma}(t) ,c^\dag_{b\sigma'}(t') \} } ,
\end{align} 
\end{subequations}
with the Heaviside step function $\theta(t)$ that takes into account causality.
Under the assumption that the center will be later dressed by a $T$-periodic drive at driving frequency $\Omega = 2\pi/T$, we will show below that the general current in the two terminals can be expressed as a Fourier series.
For now, the bare Green's functions $g$ of the individual (uncoupled, undriven) subsystems describe systems in equilibrium and, hence, only depend on the time difference $t-t'$. 
This is generally not true for the dressed Green's functions $G$ that will later include the drive applied to the center. 
With the help of the left and right Dyson equations for the nonlocal lesser Green's functions (Langreth rules), 
\begin{subequations}
\begin{align}
	G^{<}_{aC}(t,t') 
	&= 
	\int_\mathbb{R} dt_1 
	g^{<}_{aa}(t-t_1) V_{aC} G_{CC}^{\mathrm{adv}}(t_1,t') 
	+
	\int_\mathbb{R} dt_1 
	g^{\mathrm{ret}}_{aa}(t-t_1)  V_{aC} G^{<}_{CC}(t_1,t') ,
	\\
	G^{<}_{Ca}(t,t') 
	&=
	\int_\mathbb{R} dt_1 
	G_{CC}^{<}(t,t_1) V_{Ca} g^{\mathrm{adv}}_{aa}(t_1-t') 
	+ 
	\int_\mathbb{R} dt_1 
	G_{CC}^{\mathrm{ret}}(t,t_1) V_{Ca} g^{<}_{aa}(t_1-t'), 
\end{align} 
\end{subequations}
we relate the nonlocal lesser Green's functions with the local Green's functions of the terminals and the center.

Assuming that the total Hamiltonian of the entire system is $T$-periodic after applying the drive to the center, we can already exploit the $T$-periodicity of the dressed Green's functions $G(t,t') = G(t+T,t'+T)$ at this stage.
At this point there are now a few slightly varying definitions across the literature how to exploit this time periodicity when proceeding with the nonequilibrium Green's function formalism. 
After performing a Fourier transform in the relative time $\tau = t-t'$, in which there is no periodicity of the Green's function, the main difference boils down to whether the resulting Green's function is considered as a periodic function in one of the three variables: $t$, $t'$, or the average time $(t+t')/2$; see, for instance, Refs.~\cite{Kohler2005,Liu2017,Mosallanejad2024b}.

We choose to work with the convention $G(t,t') = G(t,t-\tau)$ used in Ref.~\cite{Kohler2005}, i.e., periodicity in $t$.
Under this assumption, all dressed Green's functions can be expanded as 
\begin{align}
	G(t,t') = \sum_{n\in\mathbb{Z}} e^{i n \Omega t} \int_\mathbb{R} \frac{dE}{2\pi\hbar} e^{-iE (t-t') / \hbar}   \mathcal{G}_n(E) ,
	\label{eq:expansionFourierG}
\end{align}
with the inverse expression
\begin{align}
	\mathcal{G}_n(E)  = \frac{1}{T} \int_0^T dt \int_\mathbb{R} d\tau \, e^{i E \tau/\hbar}  e^{-i n \Omega t} G(t,t-\tau)  .
\end{align}
Using the Fourier representation of the bare lead Green's functions,
\begin{align}
	g(t-t') = \int_\mathbb{R} \frac{dE}{2\pi\hbar} e^{-iE(t-t')/\hbar} g(E),
\end{align}
we finally arrive at the Fourier series representation of  Eq.~\eqref{eq:currentStartingPoint} as 
\begin{align}
	I_a(t) &= \sum_{n\in\mathbb{Z}} I_{a,n} \, e^{in\Omega t}  ,
\end{align}
with Fourier coefficients
\begin{align}
	 I_{a,n}
	 =
	 \frac{e}{h}  \int_\mathbb{R} dE  
	\, & \mathrm{Tr}\Bigl(    
	V_{Ca} g^{<}_{aa}(E-n\hbar\Omega) V_{aC} 
	\mathcal{G}_{CC,n}^{\mathrm{adv}}(E) 
	+ 
	V_{Ca} g^{\mathrm{ret}}_{aa}(E-n\hbar\Omega) 
	V_{aC} \mathcal{G}_{CC,n}^{<}(E) 
	\nonumber \\
	&-   
	\mathcal{G}_{CC,n}^{<}(E) 
	V_{Ca} g^{\mathrm{adv}}_{aa}(E) V_{aC} 
	-  
	\mathcal{G}_{CC,n}^{\mathrm{ret}}(E) 
	V_{Ca} g^{<}_{aa}(E) V_{aC} 
	\Bigr)  .
\end{align}   
In the following, we are only interested in the time-averaged current, which is given by the Fourier coefficient $I_{a,0}$. 
Using the general relation $G^\mathrm{adv} - G^\mathrm{ret} = G^< - G^>$ for nonequilibrium Green's functions, we obtain
\begin{align}
	I_{a,0}
	=
	\frac{e}{h}  \int_\mathbb{R} dE  
	\, \mathrm{Tr}\Bigl(    
	V_{Ca} g^{>}_{aa}(E) V_{aC} 
	\mathcal{G}_{CC,0}^{<}(E) 
	-  
	V_{Ca} g^{<}_{aa}(E) V_{aC} 
	\mathcal{G}_{CC,0}^{>}(E) 
	\Bigr)  .
\end{align}   
The remaining task is to determine $\mathcal{G}_{CC,0}^{<,>}$. 
To do so, we use
\begin{align}
	G_{CC}^{<,>}(t,t')
	=
	\sum_{a=L,R} 
	\int_{\mathbb{R}} dt_1  \int_\mathbb{R} dt_2 \, G^{\mathrm{ret}}_{CC}(t,t_1) V_{Ca} g_{aa}^{<,>}(t_1-t_2) V_{aC} G_{CC}^\mathrm{adv}(t_2,t') ,
\end{align}
which translates to
\begin{align}
	 \mathcal{G}_{CC,n}^{<,>}(E) 
	 =
	  \sum_{b=L,R}  
	   \sum_{k,l \in \mathbb{Z}}   
	     \delta_{k+l,n}  \, 
	   \mathcal{G}^{\mathrm{ret}}_{CC,k}(E  - l \hbar\Omega) V_{Cb}
	  g_{bb}^{<,>}(E - l \hbar \Omega) 
	  V_{bC} \mathcal{G}^{\mathrm{adv}}_{CC,l}(E) .
\end{align}
Finally, we use the equilibrium expressions of the lesser and greater Green's functions of the terminals,
\begin{align}
	g^<_{aa}(E) = 2 \pi i \,  \rho_a(E) \, f_a(E),
	\qquad 
	g^>_{aa}(E) = -2\pi i \, \rho_a(E) \, [1-f_a(E)], 
\end{align}
with the local density of states $\rho_a(E) =  \mathrm{Im}[g^\mathrm{adv}_{aa}(E)]/\pi$ that generally takes into account the electronic structure of terminal $a$, and $f_a(E) = [1+\exp(\beta(E-\mu_a))]^{-1}$ is the Fermi function in terminal $a$ at chemical potential $\mu_a$ and inverse temperature $\beta$. 
Then, we have 
\begin{align}
	I_{a,0}
	&=
	\frac{e}{h}  
	\int_\mathbb{R} dE  
	\sum_{b=L,R}  
	\sum_{k \in \mathbb{Z}}    
	\, \mathrm{Tr}\Bigl(    
	\Gamma_a(E) 
	\mathcal{G}^{\mathrm{ret}}_{CC,k}(E + k \hbar\Omega) 
	\Gamma_b(E + k \hbar \Omega) \mathcal{G}^{\mathrm{adv}}_{CC,-k}(E)
	\Bigr) 
	\Bigl(  f_b(E + k \hbar \Omega) - f_a(E) \Bigr)
	 ,
	 \label{eq:currentBeforeSigmund}
\end{align}   
where we introduced the coupling matrices $\Gamma_a(E) = 2 \pi V_{Ca} \rho_a(E) V_{aC}$ that define the specific connections of the terminals to the center and takes into account the electronic structure of the leads.
Shifting the energy $E \to E - k \hbar \Omega$ of the term proportional to $f_b$, using the relation $\mathcal{G}_{k}^\mathrm{adv}(E) = [\mathcal{G}_{-k}^\mathrm{ret}(E-k\hbar\Omega)]^\dag$ that follows from $[G^\mathrm{adv}(t,t')]^\dag = G^{\mathrm{ret}}(t',t)$ in the time domain, and swapping the sign $k \to -k$ in the second term, we arrive at
\begin{align}
	I_{a,0}
	&=
	\sum_{b=L,R}  
	\frac{e}{h}  
	\int_\mathbb{R} dE  
	\sum_{k \in \mathbb{Z}}    
	\Bigl(
	\mathcal{T}_{k,ab}^\mathrm{ret}(E)
	f_b(E) 
	-
	\mathcal{T}_{k,ba}^\mathrm{adv}(E)
	f_a(E)
	\Bigr)
	,
	\label{eq:currentFormula}
\end{align}   
where we defined the transmission functions
\begin{subequations}
\begin{align}
	\mathcal{T}_{k,ab}^{\mathrm{ret}}(E) 
	&=
	\mathrm{Tr}\Bigl(    
	\Gamma_a(E - k \hbar \Omega) 
	\mathcal{G}^{\mathrm{ret}}_{CC,k}(E) 
	\Gamma_b(E) 
	[\mathcal{G}_{CC,k}^\mathrm{ret}(E)]^\dag
	\Bigr) ,
	\\
	\mathcal{T}_{k,ba}^\mathrm{adv}(E)
	&= 
	\mathrm{Tr}\Bigl(    
	\Gamma_b(E - k \hbar \Omega) \mathcal{G}^{\mathrm{adv}}_{CC,k}(E)
	\Gamma_a(E) 
	[\mathcal{G}_{CC,k}^\mathrm{adv}(E)]^\dag
	\Bigr) .
\end{align}
\end{subequations}
Since the dressing by the drive formally does not distinguish between retarded and advanced perturbations, the labels "retarded" and "advanced" can be dropped by restricting to one of the two types of Green's functions. 
This is also supported by numerical evaluation of the transmission functions and is due to the combination of $\mathcal{G}$ and $\mathcal{G}^\dag$ appearing in the trace. 
Therefore, $\mathcal{T}_{k,ab}^{\mathrm{ret}}(E) = \mathcal{T}_{k,ab}^{\mathrm{adv}}(E) \equiv \mathcal{T}_{ab}^{(k)}(E)$.
However, note that generally $\mathcal{T}_{ab}^{(k)}(E) \neq \mathcal{T}_{ba}^{(k)}(E)$, as discussed in detail in Ref.~\cite{Kohler2005}, and equality only holds for $k=0$. 
Finally, the fact that $\mathcal{T}_{k,ab}^{\mathrm{ret}}(E) = \mathcal{T}_{k,ab}^{\mathrm{adv}}(E)$ holds also results in $	I_{R,0} = - I_{L,0}$, which also makes sense on a physical level, since the drive in the center should not create additional particles which outflow to both terminals.

To conclude, we present $I_0 \equiv I_{R,0} = - I_{L,0}$ in Eq.~\eqref{eq:averageCurrent} in the main text, and the transmission function $\mathcal{T}_{k,ab}^{\mathrm{ret}}(E) \equiv \mathcal{T}_{ab}^{(k)}(E)$ is calculated from the retarded Green's function $\mathcal{G}^{\mathrm{ret}}_{CC,k}(E) \equiv \mathcal{G}_{k}(E)$. We also drop the label $CC$ for the center Green's function to clean up the notation for readability, which brings us to the final form \cite{Kohler2005}
\begin{align}
	I_{0}
	&= 
	\frac{e}{h}  
		\int_\mathbb{R} dE  
		\sum_{k \in \mathbb{Z}}    
		\Bigl[
		\mathcal{T}_{RL}^{(k)}(E)
		 f_L(E) 
		 - 
		 \mathcal{T}_{RL}^{(k)}(E) 
		 f_L(E) \Bigr] .
\end{align}   
As a last step, we define the average and relative transmissions
\begin{align}
	\mathcal{T}_\mathrm{avg}^{(k)}(E)
	&=
	\frac{1}{2} \Bigl[ 
	\mathcal{T}_{RL}^{(k)}(E)
	+
	\mathcal{T}_{LR}^{(k)}(E)
	\Bigr] ,
	\qquad 
	\mathcal{T}_\mathrm{rel}^{(k)}(E)
	=
	\frac{1}{2} \Bigl[ 
	\mathcal{T}_{RL}^{(k)}(E)
	-
	\mathcal{T}_{LR}^{(k)}(E)
	\Bigr] ,
\end{align}
respectively, which allows us to separate the current $I_0 = I_{\mathrm{Landauer}} + I_\mathrm{pump}$ into a Landauer-type contribution 
\begin{align}
	I_\mathrm{Landauer}
	&= 
	\frac{e}{h}  
	\int_\mathbb{R} dE  
	\sum_{k \in \mathbb{Z}}    
	\mathcal{T}_\mathrm{avg}^{(k)}(E)
	\Bigl[ f_L(E) - f_R(E) \Bigr] 
\end{align}
due to the voltage $V$ and a pump contribution 
\begin{align}
	I_\mathrm{pump}
	=
	\frac{e}{h}  
	\int_\mathbb{R} dE  
	\sum_{k \in \mathbb{Z}}    
	\mathcal{T}_\mathrm{rel}^{(k)}(E)
	\Bigl[ 
	f_L(E) 
	+
	f_R(E)
	\Bigr] 
\end{align}   
originating from the applied drive to the center.
In the absence of the drive, $A = 0$, we have $	\mathcal{T}_{ab}^{(k)} = 0$ for $k \neq 0$ and $\mathcal{T}_{RL}^{(0)} = \mathcal{T}_{LR}^{(0)}$. 
Hence, the relative transmission $\mathcal{T}_\mathrm{rel}^{(k)}$ vanishes, and so does the pump current, $I_\mathrm{pump} = 0$. 
In the absence of a voltage, $V = 0$, we have $f_L(E) = f_R(E)$ and the Landauer-type contribution vanishes, $I_\mathrm{Landauer} = 0$.

\section{Dressed Green's function of the center}
The remaining task is to determine the dressed Green's function of the center in the presence of a drive, including the coupling with the two terminals. 
Assuming that the center is now driven, $H(t) = H_C + W(t)$, as defined in the main text with $W(t)$ in the Fourier representation
\begin{align}
	W(t) = \sum_{k\in\mathbb{Z}} W_k e^{ik\Omega t},
\end{align}
the full dressed retarded/advanced center Green's function follows from the Dyson equation
\begin{align}
	G_{CC}^{\mathrm{ret,adv}}(t,t') 
	= g_{CC}^{\mathrm{ret,adv}}(t-t') 
	&+ 
	\int_{\mathbb{R}} dt_1 
	\int_{\mathbb{R}} dt_2
	\, g_{CC}^{\mathrm{ret,adv}}(t-t_1)
	\Sigma^{\mathrm{ret,adv}}_{CC}(t_1-t_2) G_{CC}^{\mathrm{ret,adv}}(t_2,t')
	\nonumber 
	\\
	&+ 
	\int_{\mathbb{R}} dt_1  
	\, g_{CC}^{\mathrm{ret,adv}}(t-t_1)
	W(t_1) G_{CC}^{\mathrm{ret,adv}}(t_1,t'),
	\label{eq:dysonEquationDrive}
\end{align}
where the self-energy 
\begin{align}
	\Sigma^{\mathrm{ret,adv}}_{CC}(t_1-t_2)
	=
	\sum_{b = L,R} V_{Cb} g_{bb}^{\mathrm{ret,adv}}(t_1-t_2) V_{bC}
\end{align}
includes the couplings to the two terminals.
Furthermore,  $g_{CC}^{\mathrm{ret,adv}}(E)= [E \pm i \eta - H_C]^{-1}$
is the bare Green's function of the isolated central system without the drive, where $\eta \to 0^+$ is a small Dynes parameter taking into account the causality boundary condition of the retarded and advanced Green's function [i.e., the Heaviside step function $\theta(t)$ in Eq.~\eqref{eq:retardedAdvancedDefinitions}].
By means of the expansion defined in Eq.~\eqref{eq:expansionFourierG}, Eq.~\eqref{eq:dysonEquationDrive} translates to the Floquet-Dyson equation
\begin{align}
	\mathcal{G}_{CC,n}^{\mathrm{ret,adv}}(E)
	= 
	\delta_{n0} \, g_{CC}^{\mathrm{ret,adv}}(E)
	&+
	g_{CC}^{\mathrm{ret,adv}}(E-n\hbar\Omega) 
	\Sigma^{\mathrm{ret,adv}}_{CC}(E-n\hbar\Omega)
	\mathcal{G}_{CC,n}^{\mathrm{ret,adv}}(E)
	\nonumber \\
	& +
	g_{CC}^{\mathrm{ret,adv}}(E-n\hbar \Omega)
	\sum_{k,l \in\mathbb{Z}} \delta_{k+l,n}
	W_{k} 
	\mathcal{G}_{CC,l}^{\mathrm{ret,adv}}(E)
	\label{eq:FloquetDysonEquation}
\end{align}
which enters Eq.~\eqref{eq:currentFormula}.
Finally, we can think of the process of dressing the Green's functions in two steps. The first dressing is due to the coupling with the terminals in the sense of standard transport calculations through the undriven center. The second dressing is the drive applied to the center. To make this two-step dressing process explicit, note that, after a few algebraic manipulations, Eq.~\eqref{eq:FloquetDysonEquation} is equivalent to
\begin{align}
	\label{eq:DysonEquationIncludingBroadening}
	\mathcal{G}_{CC,n}^{\mathrm{ret,adv}}(E)
	= 
	\delta_{n0}
	g_{CC,n}^{\mathrm{ret,adv}}(E)
	& +
	g_{CC,n}^{\mathrm{ret,adv}}(E)
	\sum_{k,l \in\mathbb{Z}} \delta_{k+l,n}
	W_{k} 
	\mathcal{G}_{CC,l}^{\mathrm{ret,adv}}(E) ,
\end{align}
where we defined the Green's function
\begin{align}
	g_{CC,n}^{\mathrm{ret,adv}}(E) = 
	\Bigl[
	E - n\hbar\Omega
	\pm i \eta 
	- H_C
	-
	\Sigma^{\mathrm{ret,adv}}_{CC}(E-n\hbar\Omega)
	\Bigr]^{-1}
	\label{eq:bareCenterDressedLeads}
\end{align}
that is already dressed by the coupling with the leads.
In the absence of the drive, $W_k = 0$ for all $k$, which means $H(t) = H_C$, we are left with
$\mathcal{G}_{CC,0}^{\mathrm{ret,adv}}(E) =  
g_{CC,0}^{\mathrm{ret,adv}}(E)$ and $\mathcal{G}_{CC,n}^{\mathrm{ret,adv}}(E) = 0$ for $n \neq 0$. 
This reduces to the standard textbook result for resonant tunneling through a static barrier \cite{cuevas2nd}.

Finally, a common approximation in electronic transport setups is the wide-band limit, in which the bare Green's function of the terminals is approximately constant in energy around the Fermi energy.
In this limit, both the couplings to the terminals, $\Gamma_a(E) \equiv \Gamma_a$, and  the self-energy, $\Sigma^{\mathrm{ret,adv}}_{CC}(E) \equiv \mp i (\Gamma_L + \Gamma_R)/2$, become energy-independent \cite{cuevas2nd}. 
This is the case considered in the main text, where again all Green's functions are retarded and the bare center includes already the coupling with the terminals, as defined in Eq.~\eqref{eq:bareCenterDressedLeads}.

\section{Linear conductance and quantum geometry}
\subsection{Low-amplitude expansion and secular approximation of the transmission function}
In order to draw the connection with quantum geometry below, 
it is most useful to write the expressions $W_k$ themselves in a perturbative series in powers of a small driving amplitude $A \ll 1$ including derivatives of the center Hamiltonian $H_C$.
We consider the simultaneous two-parameter drive of the form $\lambda_\alpha(t) = \lambda_\alpha + A \cos(\Omega t)$ and $\lambda_\beta(t) = \lambda_\beta + A \cos(\Omega t-\gamma)$ with a relative phase difference $\gamma$ and $\alpha \neq \beta$. 
Then, $H(t) = H_C + W(t)$, where $\partial_\alpha = \partial / \partial \lambda_\alpha$ and 
\begin{subequations}
	\begin{align} 
		W(t) &=  
		\sum_{k=-2}^2 W_k e^{i k \Omega t} + \mathcal{O}(A^3), 
		\\
		W_0 &= 
		\frac{1}{4} A^2 \Bigl[ 
		\partial_\alpha^2 H_C
		+ \partial_\beta^2 H_C
		+ 2 \cos(\gamma) \, \partial_\alpha \partial_\beta H_C
		\Bigr] , 
		\\
		\label{eq:Wpm1}
		W_{\pm 1} &=
		\frac{1}{2} A \Bigl[ \partial_\alpha H_C + e^{\mp i\gamma}\partial_\beta H_C \Bigr] ,
		\\
		W_{\pm 2} &= 
		\frac{1}{8} A^2 \Bigl[
		\partial_\alpha^2 H_C
		+ 2 e^{\mp i \gamma} \partial_\alpha \partial_\beta H_C
		+ e^{\mp 2 i \gamma}  \partial_\beta^2 H_C
		\Bigr]
		.
	\end{align}
\end{subequations}
The case of single-parameter drives, $\lambda_\alpha(t) = \lambda_\alpha + A \cos(\Omega t)$, follows from dropping all terms above that involve derivatives $\partial_\beta$. 
By using the Dyson equation \eqref{eq:DysonEquationIncludingBroadening} repeatedly, the relevant terms for the full transmission function are
\begin{subequations}
	\begin{align}  
		\mathcal{T}_{ab}^{(0)}
		=
		\mathrm{Tr}\left(    
		\Gamma_a
		\mathcal{G}_{0}
		\Gamma_b
		\mathcal{G}_{0}^\dag
		\right)
		&=
		\mathrm{Tr}\left(    
		\Gamma_a
		g_0
		\Gamma_b
		g_0^\dag
		\right)
		+
		\mathrm{Tr}\left(    
		\Gamma_a
		g_0
		\Gamma_b 
		g_0^\dag W_{0} g_{0}^\dag
		\right)
		+ 
		\mathrm{Tr}\left(    
		\Gamma_a
		g_0
		W_{0} 
		g_{0} 
		\Gamma_b
		g_0^\dag
		\right)
		\nonumber \\
		&
		\quad + 
		\mathrm{Tr}\left(    
		\Gamma_a
		g_0
		\Gamma_b 
		g_0^\dag
		W_{1} 
		g_{-1}^\dag 
		W_{-1} 
		g_{0}^\dag
		\right)
		+ 
		\mathrm{Tr}\left(    
		\Gamma_a
		g_0
		\Gamma_b 
		g_0^\dag W_{-1} g_1^\dag
		W_{1} 
		g_{0}^\dag
		\right)
		\nonumber \\
		&
		\quad + 
		\mathrm{Tr}\left(    
		\Gamma_a
		g_0
		W_{1} 
		g_{-1} 
		W_{-1} 
		g_{0}
		\Gamma_b
		g_0^\dag
		\right)
		+ 
		\mathrm{Tr}\left(    
		\Gamma_a
		g_0
		W_{-1} 
		g_1
		W_{1} 
		g_{0}
		\Gamma_b
		g_0^\dag
		\right)
		+
		\mathcal{O}(A^3) ,
		\\
		\mathcal{T}_{ab}^{(\pm 1)}
		=
		\mathrm{Tr}\left(    
		\Gamma_a
		\mathcal{G}_{\pm 1}
		\Gamma_b
		\mathcal{G}_{\pm 1}^\dag
		\right)
		&=
		\mathrm{Tr}\left(    
		\Gamma_a
		g_{\pm 1}
		W_{\pm 1} 
		g_{0}
		\Gamma_b
		g_0^\dag
		W_{\pm 1}^\dag
		g_{\pm 1}^\dag
		\right) 
		+
		\mathcal{O}(A^3) 
		\\
		\mathcal{T}_{ab}^{(k)} &= \mathcal{O}(A^{2 |k|})  \qquad (|k| \geq 2)
	\end{align}
\end{subequations}
Note that we have used the general relation $W_k^\dag = W_{-k}$, which follows from the hermicity of $W(t)$, and that terms with $W_{\pm2} \propto A^2$ are not relevant, since they will only contribute to higher orders in the transmission functions. 
This means that the lowest-order correction to the current will be $\propto A^2$ and that we only have to consider terms with $k = -1,0,1$ for the transmission functions $\mathcal{T}_{ab}^{(k)}$. 
We can also identify the part $\mathrm{Tr}(\Gamma_a g_0 \Gamma_b g_0^\dag)$ of the transmission function $\mathcal{T}_{ab}^{(0)}$ that describes the resonant tunneling in the absence of the drive (i.e., $W_k=0$ for $A=0$) \cite{cuevas2nd}.

We evaluate the trace in the eigenstates $\ket{n}$ of the bare Hamiltonian $H_C$. 
For generally nonuniform couplings $\Gamma_a$ of the center to the terminals, both the dressed Green's function in Eq.~\eqref{eq:bareCenterDressedLeads} and the coupling matrices expressed in the eigenstates will not be diagonal.
$g_{k} = N / D$ can generally be written with a numerator matrix $N$ and a common denominator function $D$. 
In the tunneling regime, a very good analytical approximation of the common denominator is given by the product 
\begin{align}
	D \approx \prod_n ( E-k\hbar \Omega - \varepsilon_n + i (\Gamma_{L,n} + \Gamma_{R,n})/2 ) ,
\end{align}
with $\Gamma_{a,n} = \braket{n|\Gamma_a|n}$.
The off-diagonal elements of the numerator matrix $N$ are already of the order $\mathcal{O}(\Gamma_a)$ and can therefore be neglected in the tunneling regime.
This allows us to approximate the full solution by the Lehmann representation
\begin{align}
	\label{eq:LehmannApproximate}
	g_k(E) \approx \sum_n \frac{ \ket{n}\bra{n} }{
		E- k \hbar \Omega  - \varepsilon_n + i (\Gamma_{L,n} + \Gamma_{R,n})/2} ,
\end{align}
which is also presented in the End Matter, Appendix B, in Eq.~\eqref{eq:apprixLehmann}.
As derived in the End Matter, Appendix A, of the main text in Eq.~\eqref{eq:linearConductance}, the linear conductance $G$ depends on the average transmission functions $\mathcal{T}_{\mathrm{avg}}^{(k)} = (\mathcal{T}_{RL}^{(k)} + \mathcal{T}_{LR}^{(k)} )/2$.
Even though the trace formulas for the transmission functions explicitly contain the total matrices $\braket{m|\Gamma_a|n}$, restricting the trace only to the diagonal elements $\braket{m|\Gamma_a|m}$ represents a valid approximation if the center is driven resonantly (secular approximation) in the tunneling regime and for spectral energies around the bound state energies of the center.
Under these approximations, the corrections $\propto A^2$ of the transmission functions are
\begin{subequations}
\begin{align}
	\delta\mathcal{T}_{\mathrm{avg}}^{(0)}
	&\approx
	2 \sum_m 
	\Gamma_{R,m} \Gamma_{L,m} 
	\left|\braket{m|g_0|m}\right|^2
	\biggl\{ 
	\braket{m|W_{0}|m} 
	\mathrm{Re}\Bigl[ \braket{m|g_0|m}  \Bigr]
	\nonumber \\
	&
	\quad +
	\sum_{n \atop (n\neq m)}
	\left(
	\left|\braket{n|W_{-1}|m}\right|^2
	\mathrm{Re}\bigl[ 
	\braket{m|g_0|m}
	\braket{n|g_{-1}|n}
	\bigr]
	+ 
	\left|\braket{n|W_{+1}|m}\right|^2
	\mathrm{Re}\bigl[ 
	\braket{m|g_0|m}
	\braket{n|g_{+1}|n}
	\bigr]
	\right)
	\biggr\} ,
	\\
	\delta\mathcal{T}_{\mathrm{avg}}^{(\pm1)}
	&\approx
	\sum_{m,n\atop (m\neq n)}
	\frac{1}{2}
	\Bigl(
	\Gamma_{R,m} \Gamma_{L,n}
	+
	\Gamma_{L,m} \Gamma_{R,n}
	\Bigr)
	\left|\braket{n|g_{0}|n}\right|^2 
	\left|\braket{m|g_{\pm1}|m}\right|^2
	\left|\braket{m|W_{\pm1}|n}\right|^2 .
\end{align}
\end{subequations}
Note that $\mathcal{T}_{\mathrm{avg}}^{(0)}$ contains terms that are of the order $A^0$ (i.e., $W_k = 0$), which are the standard resonant tunneling contributions in the absence of the drive.
These are to be removed by a reference measurement of the conductance for $A = 0$ to identify the change in the conductance $\delta G \propto A^2$. 

\subsection{Change in conductance in the tunneling regime}
As explained in the main text, the goal is to measure the linear conductance change $\delta G_n = \delta G(\varepsilon_n) \propto A^2$ through an energy level via $\mu = \varepsilon_n$ for a small voltage, while driving resonantly at $\hbar \Omega = |\varepsilon_n - \varepsilon_p|$ with $p \neq n$. 
Evaluating the transmission functions at $E = \mu = \varepsilon_n$, $\hbar\Omega = | \varepsilon_n - \varepsilon_p|$ $(p\neq n)$, the dominant corrections in the tunneling limit are of the order $\mathcal{O}(\Gamma^{-2})$ and read
\begin{subequations}
\begin{align}
	\delta \mathcal{T}_{\mathrm{avg}}^{(0)}(\varepsilon_n)
	&\approx 
	\left\{ 
	\begin{array}{ccc}
		\frac{ - 32 \Gamma_{R,n} \Gamma_{L,n}  
		}{ (\Gamma_{L,n} + \Gamma_{R,n})^3 (\Gamma_{L,p} + \Gamma_{R,p}) } \left|\braket{p|W_{-1}|n}\right|^2
		& , &  \varepsilon_p >  \varepsilon_n \\
		\\
		\frac{ -32 \Gamma_{R,n} \Gamma_{L,n}   }{ (\Gamma_{L,n} + \Gamma_{R,n})^3 (\Gamma_{L,p} + \Gamma_{R,p})}
		\left|\braket{p|W_{+1}|n}\right|^2
		& , & \varepsilon_p < \varepsilon_n
	\end{array}
	\right. ,
\\
\nonumber \\
	\delta\mathcal{T}_{\mathrm{avg}}^{(-1)}(\varepsilon_n)
	&\approx  
	\left\{ 
	\begin{array}{ccc}
		\frac{ 8( \Gamma_{R,p} \Gamma_{L,n} + \Gamma_{R,n} \Gamma_{L,p} ) }{  (\Gamma_{L,n} + \Gamma_{R,n})^2   (\Gamma_{L,p} + \Gamma_{R,p})^2}
		\left|\braket{p|W_{- 1}|n}\right|^2
		& , & \varepsilon_p > \varepsilon_n \\
		\\
		0 & , & \varepsilon_p < \varepsilon_n
	\end{array}
	\right. ,
\\
\nonumber \\
	\delta\mathcal{T}_{\mathrm{avg}}^{(+1)}(\varepsilon_n)
	&\approx
	\left\{ 
	\begin{array}{ccc}
		0 & , & \varepsilon_p > \varepsilon_n \\
		\\
		\frac{ 8 (\Gamma_{R,p} \Gamma_{L,n} + \Gamma_{R,n} \Gamma_{L,p} ) }{  (\Gamma_{L,n} + \Gamma_{R,n})^2  (\Gamma_{L,p} + \Gamma_{R,p})^2}
		\left|\braket{p|W_{+ 1}|n}\right|^2
		& , & \varepsilon_p < \varepsilon_n
	\end{array}
	\right. .
\end{align} 
\end{subequations}
Hence, the dominant correction $\delta G_n \propto A^2$ to the conductance $G_n = G_n|_{A=0} + \delta G_n$ in the tunneling limit reads
\begin{align}
	\delta G_n 
	\approx 
	\frac{8 e^2}{h}  
	\left( 
	\frac{ 
		(\Gamma_{R,n} - \Gamma_{L,n}  ) (\Gamma_{R,n} \Gamma_{L,p} - \Gamma_{L,n} \Gamma_{R,p})
		- 2 \Gamma_{R,n} \Gamma_{L,n} (\Gamma_{L,p} + \Gamma_{R,p})
	}{ 
		(\Gamma_{L,n} + \Gamma_{R,n})^3 (\Gamma_{L,p} + \Gamma_{R,p})^2
	} 
	\right) 
	\left|\braket{p|W_{\sigma_{np}}|n}\right|^2 ,
\end{align}
with $\sigma_{np} = \mathrm{sgn}(\varepsilon_n - \varepsilon_p)$. 

\subsection{Relation with quantum geometry}
The relation with quantum geometry for the above mentioned two-parameter drive is drawn by noting that 
\begin{align}
	\forall p \neq n: 
	\quad \braket{n|\partial_\alpha H_C|p}
	\braket{p|\partial_\beta H_C|n} &=  (\varepsilon_p - \varepsilon_n)^2 q_{\alpha\beta,pn} ,
\end{align}
where the QGT elements $q_{\alpha\beta,pn}$ are defined in Eq.~\eqref{eq:qgt} in the main text.
By means of the expressions for $W_{\pm1}$ in Eq.~\eqref{eq:Wpm1}, this translates to 
\begin{align}
	\forall p \neq n: \quad &
	\left|\braket{p|W_{\sigma_{np}}|n}\right|^2
	= \frac{A^2}{4}  (\varepsilon_p - \varepsilon_n)^2
	\Bigl[ 
	q_{\alpha\alpha,pn}
	+
	q_{\beta\beta,pn}
	+
	2 \cos(\gamma) \mathrm{Re}(q_{\alpha\beta,pn})
	+ 
	2 \sigma_{np} \sin(\gamma) \mathrm{Im}(q_{\alpha\beta,pn})
	\Bigr] .
\end{align}
Hence, the change in conductance can be written as
\begin{align}
	\delta G_n 
	\approx 
	\frac{e^2}{h}  A^2  
	P_{pn} 
	\Bigl[ 
	q_{\alpha\alpha,pn}
	+
	q_{\beta\beta,pn}
	+
	2 \cos(\gamma) \mathrm{Re}(q_{\alpha\beta,pn})
	+ 
	2 \sigma_{np} \sin(\gamma) \mathrm{Im}(q_{\alpha\beta,pn})
	\Bigr] ,
\end{align}
with $\sigma_{np} = \mathrm{sgn}(\varepsilon_n - \varepsilon_p)$ and a generally parameter-dependent prefactor
\begin{align}
	P_{pn} = 
	2 
	(\varepsilon_p - \varepsilon_n)^2
	\left( 
	\frac{ 
		(\Gamma_{R,n} - \Gamma_{L,n}  ) (\Gamma_{R,n} \Gamma_{L,p} - \Gamma_{L,n} \Gamma_{R,p})
		- 2 \Gamma_{R,n} \Gamma_{L,n} (\Gamma_{L,p} + \Gamma_{R,p})
	}{ 
		(\Gamma_{L,n} + \Gamma_{R,n})^3 (\Gamma_{L,p} + \Gamma_{R,p})^2
	} 
	\right) .
\end{align}
Again, keep in mind that for the single-parameter case, all terms with label $\beta$ are ignored. 
As a last simplification, we introduce the average broadening $\Gamma_n = (\Gamma_{L,n} + \Gamma_{R,n})/2$ and the relative difference $\delta_n = (\Gamma_{L,n} - \Gamma_{R,n})/(\Gamma_{L,n} + \Gamma_{R,n})$ to arrive at 
\begin{align}
	P_{pn} = 
	-  
	\frac{ 
		(\varepsilon_p - \varepsilon_n)^2
	}{ 
		4 \Gamma_n \Gamma_p
	} (1  + \delta_n \delta_p - 2 \delta_n^2) ,
\end{align}
which is presented in Eq.~\eqref{eq:parameterDependentPrefactors} in the main text.

Measuring at $\mu = \varepsilon_n$ and driving a single parameter $\lambda_\alpha \to \lambda_\alpha + A \cos(\Omega t)$ at resonance $\hbar \Omega = |\varepsilon_n - \varepsilon_p|$, we directly get the real-valued diagonal elements of the quantum geometric tensor, 
\begin{align}
	\delta G_n 
	\approx 
	\frac{e^2}{h} A^2 P_{pn}  \, q_{\alpha\alpha,pn}  .
\end{align}

Measuring at $\mu = \varepsilon_n$ and driving two parameters $\lambda_\alpha \to \lambda_\alpha + A \cos(\Omega t)$ and $\lambda_\beta \to \lambda_\beta + A \cos(\Omega t-\gamma)$ at resonance $\hbar \Omega = |\varepsilon_n - \varepsilon_p|$, we get the off-diagonal elements of the quantum geometric tensor from the changes in conductance $\delta G_1^{(\gamma)}$ that depend on the choice of the phase difference $\gamma$. 
Two measurements at $\gamma = 0$ and $\gamma = \pi$ (linear drives) yield the difference related with the real part of the off-diagonal elements,
\begin{align}
	\delta G_n^{(0)} - \delta G_n^{(\pi)} 
	\approx  
	4 \frac{e^2}{h}  A^2
	P_{pn}  \, 
	\mathrm{Re}(q_{\alpha\beta,pn})  .
\end{align}
Two measurements at $\gamma = \pm\pi/2$ (circular drives) yield the difference related with the imaginary part of the off-diagonal elements,
\begin{align}
	\delta G_n^{(-\pi/2)} - \delta G_n^{(+\pi/2)} 
	\approx 
	4  \, \mathrm{sgn}(\varepsilon_p - \varepsilon_n)
	\frac{e^2}{h}  A^2
	P_{pn} \, \mathrm{Im}(q_{\alpha\beta,pn}) .
\end{align}

\section{Fourier coefficients of the perturbation $W(t)$}
Let $H_C(\lambda)$ be a general center Hamiltonian parametrized by a set of parameters $\lambda = (\lambda_\alpha)$.
Under a $T$-periodic drive of the parameters, $\lambda \to \lambda(t) = \lambda(t+T)$, we write the driven Hamiltonian $H(t) = H_C(\lambda(t))$ as $H(t) = H_C + W(t)$, where $W(t) = \sum_{k \in \mathbb{Z}} W_k e^{ik\Omega t}$ is the Fourier expansion of the perturbation, with Fourier coefficients $W_k$ defined in Eq.~\eqref{eq:FourierCoefficientsMainText} in the main text. 
These Fourier coefficients are used in the numerical calculation of the conductance through the device when solving the Dyson equation \eqref{eq:DysonEquationIncludingBroadening} for the retarded Green's function elements.

In the following, we list the Fourier coefficients $W_k$ for the two-level and three-level system for different one- and two-parameter drives discussed in the main text. 
Useful relations to arrive at the final expressions are the well-known Jacobi-Anger expansion
\begin{align}
	e^{i A \cos(\Omega t)} = \sum_{k\in\mathbb{Z}} i^k J_k(A)  \,  e^{i k \Omega t} ,
\end{align}
where $J_n(z)$ are the Bessel functions of the first kind, and the less-known identity \cite{Levine2022}
\begin{align}
	\sum_{l\in\mathbb{Z}}  (\pm i)^l  J_l(A) J_{k-l}(A) 
	=  J_{k}(\sqrt{2} A) \,  e^{\pm i   k \pi / 4}  .
\end{align}

\subsection{Example 1: Two-level system}
The Hamiltonian $H_C$ of the two-level system is defined in the End Matter of the main text in Eq.~\eqref{eq:HamiltonianTwoLevelSystem} and the two parameters of the system are $(\lambda_1,\lambda_2) = (\theta,\phi)$. 
Driving only $\theta \to \theta + A \cos(\Omega t)$, we have
\begin{align}
	W_k =
	\begin{pmatrix}
		B \cos(\theta + k\pi/2)  [J_k(A) - \delta_{k0}] 
		& 
		B e^{-i\phi}
		\sin(\theta + k\pi/2)  [J_k(A) - \delta_{k0}]
		\\
		B e^{i\phi} \sin(\theta + k\pi/2)  [J_k(A) - \delta_{k0}]
		& 
		-B \cos(\theta + k\pi/2)  [J_k(A) - \delta_{k0}]
		\\
	\end{pmatrix} ,
\end{align}
while driving only $\phi \to \phi + A \cos(\Omega t)$ yields
\begin{align}
	W_k = 
		\begin{pmatrix}
			0 & B \sin(\theta) e^{-i\phi} [ (-i)^k J_k(A) - \delta_{k0} ] \\
			B \sin(\theta) e^{i\phi} [ i^k J_k(A) - \delta_{k0} ] & 0 \\
		\end{pmatrix}  .
\end{align} 
The simultaneous drive of $\theta \to \theta + A \cos(\Omega t)$ and $\phi \to \phi + A \cos(\Omega t- \gamma)$ results in 
\begin{align}
	 & W_k = 
	\left\{ 
	\begin{array}{ccc}
		\begin{pmatrix}
			B \cos(\theta + k\pi/2)  [J_k(A) - \delta_{k0}]  
			& 
			i B e^{-i\theta}  e^{- i\phi} [ (-i)^k J_k(2A) - \delta_{k0} ] / 2
			\\
			-i B e^{i\theta}   e^{ i\phi} [ i^k J_k(2A) - \delta_{k0} ] /2 
			& 
			-B \cos(\theta + k\pi/2)  [J_k(A) - \delta_{k0}]  \\
		\end{pmatrix}  
		& , & \gamma = 0 
		\\
		\\
		\begin{pmatrix}
			B \cos(\theta + k\pi/2)  [J_k(A) - \delta_{k0}]  & 
			-i B e^{i\theta}  e^{- i\phi} 
			[ i^k J_k(2A) -\delta_{k0} ] / 2 \\
			i B e^{-i\theta} e^{ i\phi} 
			[ (-i)^k J_k(2A)   -  \delta_{k0} ]   /2
			& -B \cos(\theta + k\pi/2)  [J_k(A) - \delta_{k0}]  \\
		\end{pmatrix} 
		& , & \gamma = \pi 
		\\
		\\
		\begin{pmatrix}
			B \cos(\theta + k\pi/2)  [J_k(A) - \delta_{k0}]  & B e^{- i\phi} [ \sin(\theta-   k\pi  / 4)  J_{-k}(\sqrt{2} A)  - \delta_{k0}    \sin(\theta) ] \\
			B e^{ i\phi} [  \sin(\theta+ k \pi / 4)   J_{k}(\sqrt{2} A) 
			- \delta_{k0}  \sin(\theta)  ] & -B \cos(\theta + k\pi/2)  [J_k(A) - \delta_{k0}]  \\
		\end{pmatrix} 
		& , & \gamma = \frac{\pi}{2}
		\\
		\\
		\begin{pmatrix}
			B \cos(\theta + k\pi/2)  [J_k(A) - \delta_{k0}]  & B e^{- i\phi} [
			\sin(\theta +  k \pi / 4) J_{k}(\sqrt{2} A) -  \delta_{k0} \sin(\theta) ] \\
			B e^{ i\phi} [  
			\sin(\theta - k \pi / 4) J_{-k}(\sqrt{2} A) - \delta_{k0}  \sin(\theta)  ] & -B \cos(\theta + k\pi/2)  [J_k(A) - \delta_{k0}]  \\
		\end{pmatrix} 
		& , & \gamma = - \frac{\pi}{2}
	\end{array}
	\right. .
\end{align}

\subsection{Example 2: Three-level system}
The Hamiltonian $H_C$ of the three-level system is defined in Eq.~\eqref{eq:HamiltonianThreeLevelSystem} and the three parameters of the system are $(\lambda_1,\lambda_2,\lambda_3) = (\theta,\phi_1,\phi_2)$. 
Driving only $\theta \to \theta + A \cos(\Omega t)$, we have
\begin{align}
	W_k = \begin{pmatrix}
		0 & B  e^{-i\phi_1}  \cos(\theta + k\pi/2)  [J_k(A) - \delta_{k0}] & 0  \\
		B  e^{i\phi_1}  \cos(\theta + k\pi/2)  [J_k(A) - \delta_{k0}]  & 0 & B e^{i\phi_2} \sin(\theta + k\pi/2)  [J_k(A) - \delta_{k0}] \\
		0 & B e^{-i\phi_2} \sin(\theta + k\pi/2)  [J_k(A) - \delta_{k0}] & 0 
	\end{pmatrix} .
\end{align}
Driving only the hopping phases, 
the drive $\phi_1 \to \phi_1 + A \cos(\Omega t)$ yields
\begin{align} 
	W_k = \begin{pmatrix}
		0 & B \cos(\theta)  e^{-i\phi_1} [ (-i)^k J_k(A) - \delta_{k0} ] & 0  \\
		B \cos(\theta) e^{i\phi_1} [ i^k J_k(A) - \delta_{k0} ]  & 0 & 0 \\
		0 & 0& 0 
	\end{pmatrix} ,
\end{align}
while the drive $\phi_2 \to \phi_2 + A \cos(\Omega t)$ results in 
\begin{align} 
	W_k = \begin{pmatrix}
		0 & 0 & 0  \\
		0 & 0 & B \sin(\theta) e^{i\phi_2} [ i^k J_k(A) - \delta_{k0} ] \\
		0 & B \sin(\theta) e^{-i\phi_2} [ (-i)^k J_k(A) - \delta_{k0} ] & 0 
	\end{pmatrix} .
\end{align}
For driving two parameters simultaneously, we have three possible combinations. 
Driving $\theta \to \theta + A \cos(\Omega t)$ and $\phi_1 \to \phi_1 + A \cos(\Omega t- \gamma)$ gives
\begin{align}
	W_k &=
	\left\{ 
	\begin{array}{ccc}
		{\tiny
		\begin{pmatrix}
			0 &  B e^{-i\theta}  e^{- i\phi_1} [ (-i)^k J_k(2A) - \delta_{k0}] / 2  & 0  \\
			B e^{i\theta} e^{ i\phi_1} [ i^k J_k(2 A) - \delta_{k0} ] / 2 & 0 & B e^{i\phi_2} \sin(\theta + k\pi/2)  [J_k(A) - \delta_{k0}] \\
			0 & B e^{-i\phi_2} \sin(\theta + k\pi/2)  [J_k(A) - \delta_{k0}] & 0 
		\end{pmatrix} 
		}
		& , & \gamma = 0
		\\
		\\
		{\tiny
		\begin{pmatrix}
			0 & B  e^{i\theta}  e^{- i\phi_1} 
			[ i^k J_k(2A) - \delta_{k0}   ] / 2 & 0  \\
			B e^{-i\theta} e^{ i\phi_1} [ (-i)^k J_k(2A)  - \delta_{k0} ] / 2  & 0 & B e^{i\phi_2} \sin(\theta + k\pi/2)  [J_k(A) - \delta_{k0}] \\
			0 & B e^{-i\phi_2} \sin(\theta + k\pi/2)  [J_k(A) - \delta_{k0}] & 0 
		\end{pmatrix} 
		}
		& , & \gamma = \pi
		\\
		\\
		{\tiny
			\begin{pmatrix}
				0 & B e^{- i\phi_1}  [ \cos(\theta - k \pi / 4)   J_{-k}(\sqrt{2} A) - \delta_{k0}  \cos(\theta) ]  & 0  \\
				B  e^{ i\phi_1} [ \cos(\theta + k \pi / 4) J_{k}(\sqrt{2} A) - \delta_{k0}  \cos(\theta) ] & 0 & B e^{i\phi_2} \sin(\theta + k\pi/2)  [J_k(A) - \delta_{k0}] \\
				0 & B e^{-i\phi_2} \sin(\theta + k\pi/2)  [J_k(A) - \delta_{k0}] & 0 
			\end{pmatrix} 
		}
		& , & \gamma = \frac{\pi}{2}
		\\
		\\
		{\tiny
			\begin{pmatrix}
				0 & B e^{- i\phi_1} [ \cos(\theta + k \pi / 4)  J_{k}(\sqrt{2} A)  - \delta_{k0}  \cos(\theta) ] & 0  \\
				B  e^{ i\phi_1} [ \cos(\theta - k \pi / 4) J_{-k}(\sqrt{2} A) - \delta_{k0}  \cos(\theta) ] & 0 & B e^{i\phi_2} \sin(\theta + k\pi/2)  [J_k(A) - \delta_{k0}] \\
				0 & B e^{-i\phi_2} \sin(\theta + k\pi/2)  [J_k(A) - \delta_{k0}] & 0 
			\end{pmatrix} 
		}
		& , & \gamma = - \frac{\pi}{2}
	\end{array}
	\right. .
\end{align}
Driving $\theta \to \theta + A \cos(\Omega t)$ and $\phi_2 \to \phi_2 + A \cos(\Omega t- \gamma)$, we have
\begin{align}
	W_k &=
	\left\{ 
	\begin{array}{ccc}
		{\tiny
		\begin{pmatrix}
			0 & B  e^{-i\phi_1}  \cos(\theta + k\pi/2)  [J_k(A) - \delta_{k0}] & 0  \\
			B  e^{i\phi_1}  \cos(\theta + k\pi/2)  [J_k(A) - \delta_{k0}]  & 0& 
			- i B  e^{i\theta}  e^{ i\phi_2} 
			[  i^k J_k(2A)  - \delta_{k0} ] / 2
			\\
			0 & i B  e^{-i\theta}  e^{- i\phi_2} 
			[ (-i)^k J_k(2A)  - \delta_{k0}  ] / 2 & 0 
		\end{pmatrix} 
		}
		& , & \gamma = 0
		\\
		\\
		{\tiny
		\begin{pmatrix}
			0 & B  e^{-i\phi_1}  \cos(\theta + k\pi/2)  [J_k(A) - \delta_{k0}] & 0  \\
			B  e^{i\phi_1}  \cos(\theta + k\pi/2)  [J_k(A) - \delta_{k0}]  & 0&
			i B e^{-i\theta}  e^{ i\phi_2} [ (-i)^k J_k(2A)   - \delta_{k0}   ] / 2  \\
			0 & -i B e^{i\theta}   e^{- i\phi_2}  [ i^k J_k(2A) - \delta_{k0} ] / 2 & 0 
		\end{pmatrix} 
		}
		& , & \gamma = \pi
		\\
		\\
		{\tiny
		\begin{pmatrix}
			0 & B  e^{-i\phi_1}  \cos(\theta + k\pi/2)  [J_k(A) - \delta_{k0}] & 0  \\
			B  e^{i\phi_1}  \cos(\theta + k\pi/2)  [J_k(A) - \delta_{k0}]  & 0& B  e^{ i\phi_2}  [ \sin(\theta + k \pi / 4)   J_{k}(\sqrt{2} A) - \delta_{k0} \sin(\theta) ] \\
			0 & B e^{- i\phi_2} [ \sin(\theta-k \pi / 4) J_{-k}(\sqrt{2} A) - \delta_{k0} \sin(\theta) ] & 0 
		\end{pmatrix} 
		}
		& , & \gamma = \frac{\pi}{2}
		\\
		\\
		{\tiny
		\begin{pmatrix}
			0 & B  e^{-i\phi_1}  \cos(\theta + k\pi/2)  [J_k(A) - \delta_{k0}] & 0  \\
			B e^{i\phi_1}  \cos(\theta + k\pi/2)  [J_k(A) - \delta_{k0}]  & 0 & 
			B e^{ i\phi_2} [ \sin(\theta - k \pi / 4 )  J_{-k}(\sqrt{2} A)  - \delta_{k0} \sin(\theta) ] \\
			0 & B   e^{- i\phi_2} [\sin(\theta + k \pi / 4)    J_{k}(\sqrt{2} A) - \delta_{k0} \sin(\theta) ]  & 0 
		\end{pmatrix} 
		}
		& , & \gamma = - \frac{\pi}{2}
	\end{array}
	\right. ,
\end{align}
while driving $\phi_1 \to \phi_1 + A \cos(\Omega t)$ and $\phi_2 \to \phi_2 + A \cos(\Omega t- \gamma)$, we get
\begin{align} 
	W_k = \begin{pmatrix}
		0 & B \cos(\theta) e^{-i \phi_1 } [ (-i)^k J_k(A)   - \delta_{k0} ] & 0  \\
		B \cos(\theta) e^{i \phi_1} [ i^k J_k(A)  - \delta_{k0} ]  & 0 & 
		B \sin(\theta) e^{i \phi_2} [ i^k e^{-i k \gamma} J_k(A)  - \delta_{k0}  ] \\
		0 & B \sin(\theta) e^{-i \phi_2 } [ (-i)^k e^{-ik\gamma} J_k(A)   - \delta_{k0} ] & 0 
	\end{pmatrix} .
\end{align}

\clearpage
\end{widetext}

\end{document}